\documentclass[a4paper,12pt]{article}

\usepackage[centertags]{amsmath}
\usepackage{amssymb}

\usepackage{graphicx}
\usepackage{epsfig}
\usepackage{ulem}
\usepackage[english]{babel}
\usepackage{array}
\usepackage{amsthm}
\usepackage{latexsym}
\usepackage{xcolor}

\usepackage{yfonts}
\usepackage{mathrsfs}
\usepackage[mathcal]{euscript}

\pdfoutput=1
\usepackage{epsfig}
 \usepackage{jheppubm}
\usepackage{mathdots}
\usepackage{MnSymbol}
\usepackage{multirow}

\newcommand{\nn}{\nonumber}
\newcommand{\be}{\begin{equation}}
\newcommand{\ee}{\end{equation}}
\newcommand{\bea}{\begin{eqnarray}}
\newcommand{\eea}{\end{eqnarray}}

\newcommand{\cI}{{\cal I}}

\newcommand{\cO}{{\cal O}}

\newcommand{\cV}{\mathcal{V}}

\title{Entanglement entropy of near-extremal black hole
}

\author{I. Aref'eva, T. Rusalev and I. Volovich}
\affiliation{Steklov Mathematical Institute, Russian Academy of Sciences,\\Gubkina str. 8, 119991, Moscow, Russia}

\emailAdd{arefeva@mi-ras.ru}\emailAdd{rusalev@mi-ras.ru}\emailAdd{volovich@mi-ras.ru}
\abstract{  We study how the entanglement entropy of the Hawking radiation derived using island recipe for  the Reissner-Nordström black hole  behaves as the black hole mass decreases. A general answer to the   question essentially depends not only on the character of decreasing  of the mass but also on  decreasing  of the  charge. 
We assume the specific relationship between the charge and mass 
$Q^2=GM^2[1-\left(\frac{M}{\mu}\right)^{2\nu} ]$, which we call the constraint equation.  We discuss whether it is possible to have a constraint so that the  entanglement entropy does not have an explosion at the end of evaporation, as happens in the case of thermodynamic entropy and the entanglement entropy for the Schwarzschild black hole.  We show that for some special scaling parameters, the entanglement entropy of radiation does not explode as long as the mass of the evaporating black hole exceeds the Planck mass.

}

\begin{document}

\maketitle
\flushbottom

%%%%%%%%%%%%%%%%%%%%%%%%%%%%%%%%%%%%%%%%%%%%%%%%%%%%

\section{Introduction}

 The entropy of Hawking radiation of black holes grows up to infinity during evaporation and it is a manifestation of the information paradox \cite{Haw1,Haw2}. 
The unitarity of quantum mechanics in the presence of gravity would have to make Hawking radiation non-thermal and  the entropy of an evaporating black hole would have to decrease at late time, consistent with Page's hypothetical behavior
\cite{Page:1993wv,Page:2013dx}.  An important recent progress towards solving the black hole information paradox is the discovery of entanglement island \cite{Penington:2019npb,Almheiri:2019psf,Almheiri:2020cfm}. The ``island formula''
for the entanglement entropy of Hawking radiation is based on quantum extremal surfaces \cite{Engelhardt:2014gca}. Although the prescription of the quantum extremal surface was  proposed in the
framework of holography \cite{Ryu:2006bv,Hubeny:2007xt}, it is believed that  the island rule is applicable to black holes in
more general theories. The black holes in four and higher dimensional asymptotically flat spacetime have been considered in \cite{Almheiri:2019psy, Krishnan:2020oun, Hashimoto:2020cas, Alishahiha:2020qza, Matsuo:2020ypv, Wang:2021woy, Kim:2021gzd}.
\\

Using the island recipe for an asymptotically flat Schwarzschild eternal black hole in four spacetime dimensions, saturation at a late time of the entanglement entropy to a value that  contains the surface term of the black hole entropy, and also several additional terms has been found  in \cite{Hashimoto:2020cas}. One of these two terms depends on the logarithm of the black hole's mass, while the other is inversely proportional to the black hole's mass.
It has been noted in \cite{Arefeva:2021kfx}, the last term, under assumption of slow evaporation of black hole,  makes
 the entropy  to grow  when mass of the black hole decreases. Note also that we do not have in our disposal an exact result for the entanglement entropy of evaporating black hole, but we can use the entanglement entropy of Hawking radiation  for eternal black hole with the same mass  as a first  approximation. Or in other words  we use the  expression for entanglement entropy  obtained for the case of an eternal black hole \cite{Hashimoto:2020cas} also to the case of a slowly varying black hole mass \cite{Arefeva:2021kfx}.
\\

In this article, we address the question of how the entanglement entropy of Hawking radiation of the
Reissner-Nordström  black hole  behaves as the black hole mass decreases. The entanglement entropy of Hawking radiation of Reissner-Nordström eternal black hole  has been obtained in \cite{Kim:2021gzd, Wang:2021woy}. 
Here, as in the case of the Schwarzschild black hole \cite{Arefeva:2021kfx}, we assume that the same formula is also true for the slowly varying mass  and  charge  of  the  Reissner-Nordström  black  hole.   A general answer to the above  question essentially depends not only on the character of decreasing  of the mass but also on  decreasing  of the  charge. 
\\

The time evolution of mass $M$ and charge $Q$ during evaporation of Reissner-Nordström black hole is a subject of numerous consideration \cite{Gibbons-1875,Zaumen,Carter,Damour,
Page-1976,
Hiscock, Gabriel-2000,Sorkin:2001hf,
Ong:2019vnv}. If we accept the scenario of the evaporation of a charged black hole, when the charge disappears before the mass of the black hole becomes small, then at the end of evaporation this case reduces to the pure Schwarzschild one. Thus, at small mass an explosion of the entanglement entropy occurs \cite{Arefeva:2021kfx}. If we accept the scenario that by some reason at the end of evaporation a near-extremal case $Q \approx \sqrt{G} M$ is realized, then the temperature of the black hole and the entanglement entropy with an island, generally speaking, may not increase.
 Their behavior essentially depends on the specific relationship between the charge and mass $Q=Q(M)$ of the black hole, which we call the constraint equation. In particular, we discuss whether it is possible to have a constraint so that the  entanglement entropy does not have an explosion at the end of evaporation, as happens in the case of thermodynamic entropy \cite{Arefeva:2022guf}. We consider a special class of constraint equations
 \cite{Arefeva:2022guf} 
\bea\label{stateequation}
Q^2 = G M^2 \left[1-\left(\frac{M}{\mu}\right)^{2\nu} \right], 
\eea
where $\nu \geq 0$ is a dimensionless parameter, and $\mu > M$ has the dimension of mass and  $\left(\frac{M}{\mu}\right)^{2\nu} \ll 1$ corresponds to the
 near-extremal case.
\\

We analyze a sign of the  mass derivative of the entanglement entropy with the island under constraint equation (\ref{stateequation}). 
If the mass derivative of the entropy is positive, then the entropy decreases with time, and vice versa. It is shown that the  mass derivative of the entanglement entropy of configuration with an island is positive for $\cV_1 < \nu <\cV_2$, where  $1<\cV_1$ and $\cV_2<2$, for any  physically  meaningful  values  of parameters.
At $\nu <\cV_1$ and $\nu >\cV_2$ depending on the parameters there are cases when the entanglement entropy with an island increases with mass decreasing.
\\

Under constraint equation (\ref{stateequation}) we consider the time differential equation \cite{Hiscock} for the mass and find the time evolution of the mass $M=M(t)$ and charge $Q=Q(t)$ of the black hole.
The entanglement entropy of  configuration  without  an  island  has explicit dependence on time \cite{Kim:2021gzd, Wang:2021woy}, so knowledge of the mass evolution makes it possible to compare the entropy with and without island and find  which configuration dominates at given moment of time. For $\nu<\cV_1$  the entropies with and without an island can intersect at the moment when the entropy with the island decreases (Page time) or increases (anti-Page time), which is similar to the situation for the Schwarzschild black holes \cite{Arefeva:2021kfx}.
\\

The paper is organized as follows. In Section \ref{setup} we outline the main steps and approximations used in deriving the entanglement entropy  for the Reissner-Nordström black hole.
 In Section \ref{analysis} we consider the behavior of  entanglement entropy with island and  its derivative in respect to the mass  of a charged black hole  under the constraint equation (\ref{stateequation}). To find the behavior for small mass it is continent to study  the asymptotic behavior at $M \to 0$ of entropy with an island and temperature of a charged black hole. In section \ref{timeevolution} we study the time evolution of the mass and charge of a black hole and compare the entropy with and without the island configuration.
In Appendix \ref{appendixapproximation} we study in more details restrictions related with the  approximations used in Section \ref{analysis}.

\section{Setup}\label{setup}

The island formula for the generalized entropy is given by
\be
\label{GEformula3}
\quad S(R) 
= 
\min \left\{\mathop{\mathrm{ext}}_{\cI}\left[
\frac{\mathrm{Area}(\partial {\cI})}{4G} 
+ S_{ \rm matter}(R \cup \cI)
\right]\right\},
\ee
where $R$ is region of Hawking radiation outside the black hole, $\cI$ is the island and $\partial {\cI}$ is island boundary. Extremization on any possible island is assumed, and then taking the minimum entropy configuration.
\\

Below we suppose to apply this formula to the Reissner-Nordström black hole. 
The metric of the Reissner-Nordström black hole is given as
\bea\label{dsRN}
ds^2&=&-f(r)dt^2+f(r)^{-1}dr^2+r^2d\Omega^2, \\
f(r)&=&1-\frac{2GM}{r}+\frac{GQ^2}{r^2},
\eea
where $M$ and $Q$ are the mass and the electric charge of black hole, respectively.

\begin{figure}[h!]
	\center{\includegraphics[width=75mm]{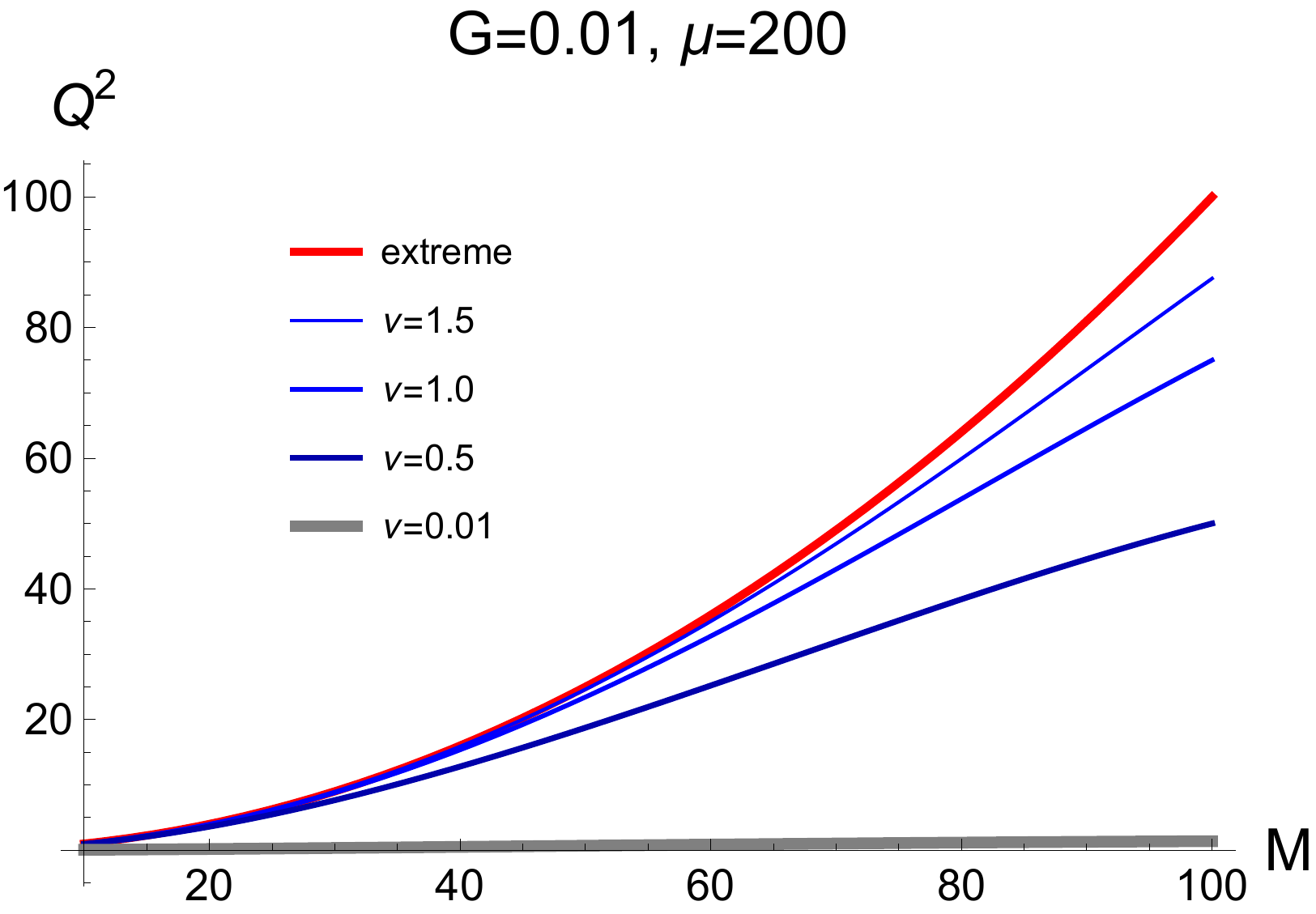}\\
	\caption{The constraint equation \eqref{stateequation} for mass $M$ and charge $Q$ of a charged black hole for different $\nu = 0.01$ (gray line) and $\nu = 0.5, 1, 1.5$ (blue lines) at fixed $\mu=200$. The red line corresponds to the extremal case $Q=\sqrt{G} M$. As $\nu$ increases, the curve $Q=Q(M)$ approaches the extremal case. The case $\nu=0.01$ is close to the Schwarzschild black hole.}} %{\bf Math.file: Graphs-Draft-Charged-BH-withMu-IA.nb}}
	\label{constraintequationfig}
\end{figure}

We are going to consider the properties of the Reissner-Nordström black hole under the constraint equation (\ref{stateequation}). The case $\nu=0$  corresponds to $Q=0$. As $\mu$ increases, the curve $Q=Q(M)$ approaches the extremal case, the same is true when $\nu$ increases at a fixed $\mu$, see Fig.\ref{constraintequationfig}. The constraint equation \eqref{stateequation} has a bell-shaped form \cite{Arefeva:2022guf} with two zeros at $M=0$ and $M=\mu$. In what  follows we consider the case $\left(\frac{M}{\mu}\right)^{2\nu} \ll 1$ that corresponds to the near-extremal case, as shown in  Fig.\ref{constraintequationfig}.

The two horizons under constraint (\ref{stateequation}) are
\be\label{horizon}
r_\pm =GM\pm\sqrt{G^2M^2-GQ^2}= GM \left[1\pm \left(\frac{M}{\mu}\right)^{\nu} \right].
\ee

The temperature of a charged black hole under constraint (\ref{stateequation}) is
\be\label{temperature}
T=\frac{1}{2\pi}\frac{\sqrt{G^2M^2-GQ^2}}{(GM+\sqrt{G^2M^2-GQ^2})^2} = \frac{M^{\nu-1}}{2 \pi G \mu^{\nu} \left(1 +  \left(\frac{M}{\mu}\right)^{\nu}\right)^2}.
\ee

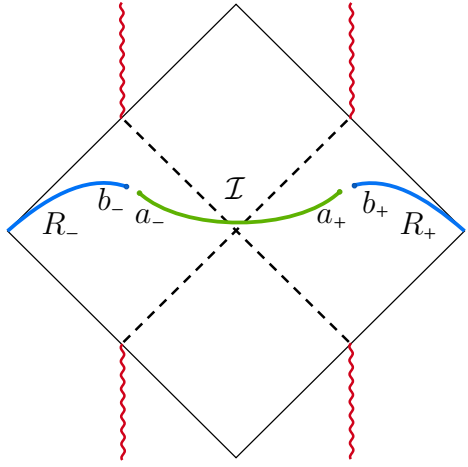
\begin{figure}[h!]
\centering
\includegraphics[width=110mm]{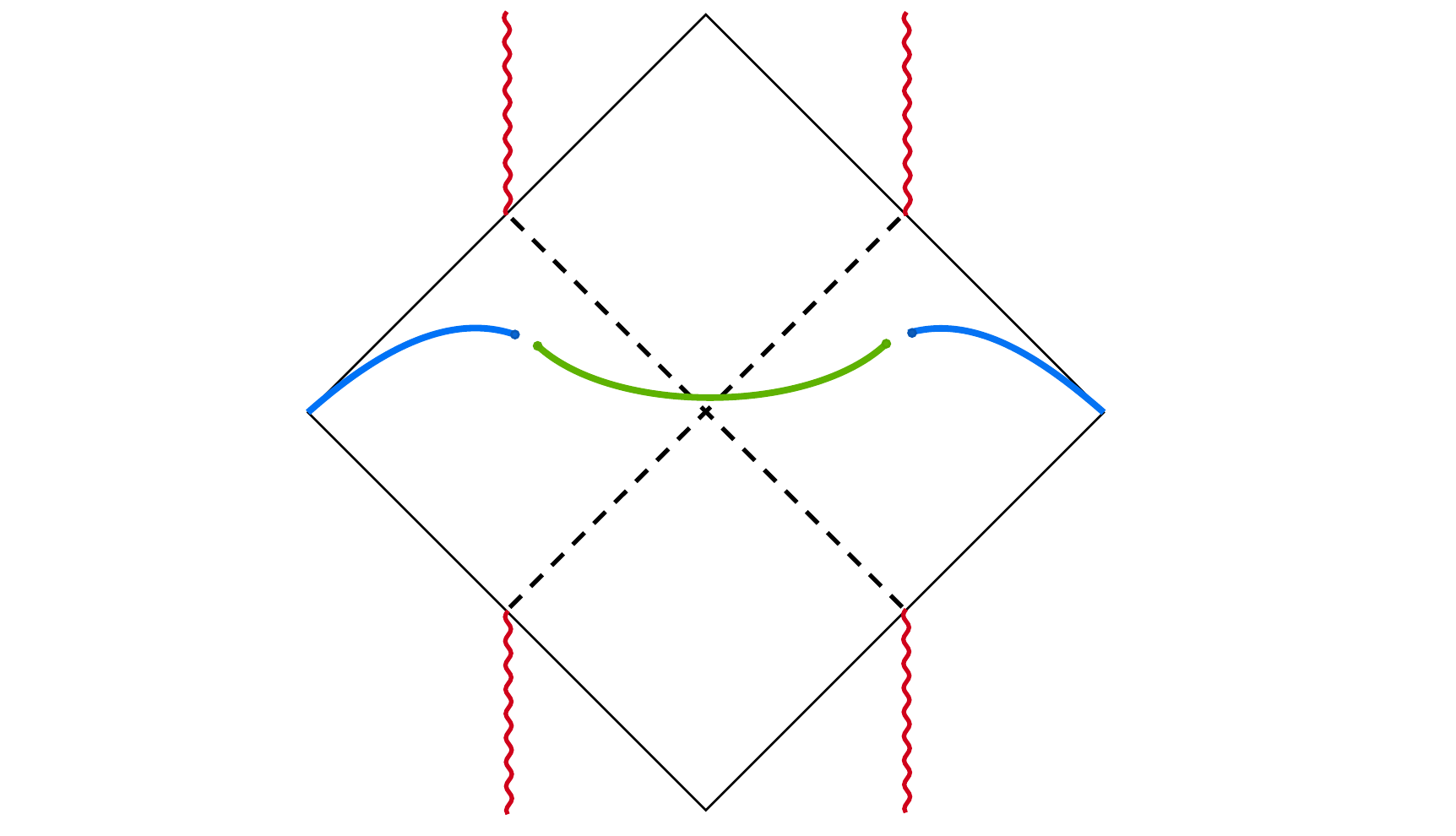}
\begin{picture}(0,0)
\put(-238,90){$\Large{R_{-}}$}
\put(-104,90){$\Large{R_{+}}$}
\put(-169,103){$\Large{\cI}$}
\put(-118,98){$\Large{b_{+}}$}
\put(-135,94){$\Large{a_{+}}$}
\put(-203,94){$\Large{a_{-}}$}
\put(-217,99){$\Large{b_{-}}$}
\end{picture}
\caption{The island configuration for the eternal non-extremal Reissner-Nordström black hole.}
	\label{fig:island}
\end{figure}

The locations of the radiation region $R_{\pm}$ and island $\cI$ are shown in Fig.\ref{fig:island}. Their boundaries are given by $b_{\pm}$ and $a_{\pm}$, respectively. Due to spherical symmetry, coordinates are defined by a pair of two  parameters $b_+ = (t_b, b)$, $b_- = (-t_b+i/2T, b)$, where $b$ is radial coordinate, similar for $a_{\pm}$. It is assumed that $b\gg r_+$ and $a=r_+ + X$, where $X > 0$, $X/r_+ \ll 1$. The former makes valid the s-wave approximation for matter part \cite{Hashimoto:2020cas} and the entropy from two-dimensional conformal field theory for matter part can be used \cite{Calabrese:2009qy}. It is also assumed that the central charge satisfies $c \gg 1$. Then effectively the radiation is described semi-classically by a two-dimensional CFT.

Under these assumptions (see Appendix \ref{appendixapproximation} for more details) the entropy with an island for asymptotically flat eternal non-extremal Reissner-Nordström black hole \cite{Kim:2021gzd, Wang:2021woy} at late times in leading order over $G$ is \footnote{Formula (\ref{islandch}) is different from formula (4.17) in \cite{Wang:2021woy} and slightly different from formula (24) in \cite{Kim:2021gzd}. The difference with \cite{Kim:2021gzd} is due to the different choice of the tortoise coordinate $r_{*} (r)$. %In \cite{Kim:2021gzd} the following coordinates are used
\bea
&&r_{*, [17]} (a) =a + \frac{r_+^2}{r_+-r_-}\log \frac{a-r_+}{r_+},\qquad r_{*, [17]} (b) =b + \frac{r_+^2}{r_+-r_-}\log \frac{b-r_+}{r_+}-\frac{r_-^2}{r_+-r_-}\log\frac{b-r_-}{r_+}, \nn\\
&& r_{*,\,here} (r) =r+ \frac{r_+^2}{r_+-r_-}\log \frac{r-r_+}{r_+}-\frac{r_-^2}{r_+-r_-}\log\frac{r-r_-}{r_-}, \, \, \, \text{for $a$ and $b$.} \nn
\eea
} 
\be\label{islandch}
	S_{\cI} = \frac{2\pi r_+^2}{G}+\frac{c}{6}\frac{(r_+ - r_-)(b-r_+)}{r_+^2} + \frac{c}{6}\log \frac{16r_+^{6}(b-r_+)^2(b-r_-)}{G^2 b^2(r_+ - r_-)^3}-\frac{c}{6}\frac{r_-^2}{r_+^2}\log\frac{b-r_-}{r_+-r_-}.
\ee

The entanglement entropy with the island \eqref{islandch} under the constraint (\ref{stateequation}) is
\bea\label{entropwithisland}\nn
S_{\cI}&=&2 \pi 
   (1+\alpha)^2 G M^2+\frac{c}{3}\frac{
   \alpha  (b-(1+\alpha) G M)}{(1+\alpha)^2 G
   M}-\frac{c}{6}\frac{(1-\alpha )^2}{(1+\alpha )^2} \log
   \left(\frac{b-(1-\alpha ) G M}{2 \alpha  G
   M}\right)\\&+&\frac{c}{6} \log \left(\frac{2 (1+\alpha)^6 G M^3 (b-(1-\alpha ) G M) (b-(1+\alpha) G M)^2}{\alpha ^3 b^2}\right),\label{entropy}\eea
   where 
   \be
   \alpha=\left(\frac{M}{\mu}\right)^{\nu}.
   \ee

One can similarly obtain the entanglement entropy of a configuration without an island \cite{Wang:2021woy, Kim:2021gzd}
\bea\label{withoutisland}
S_{n\cI} = \frac{c}{6} \log \left[\frac{16 (b-r_+)(b-r_-)r^4_+}{G b^2 (r_+-r_-)^2 } \cosh^2 \left( \frac{(r_+-r_-) t}{2 r^2_+}\right) \right].
\eea
Under the constraint \eqref{stateequation} the entropy without an island \eqref{withoutisland} is
 \bea\label{withoutislandconstraint}
S_{n\cI} &=& \frac{c}{3} \log \Big[ \frac{2 \sqrt{G} M \left(1+\alpha\right)^2 }{\alpha b} 
   \cosh \left(\frac{t  \alpha}{G M \left(1+\alpha\right)^2}\right) \\ && \sqrt{\left(b-G
   M \left(1-\alpha\right)\right) \left(b-G
   M \left(1+\alpha\right)\right)}\Big].  \nn
 \eea

 In Appendix \ref{appendixapproximation} we study in details  the approximations under which the entropy with an island \eqref{islandch} is derived.
As has been mentioned above, we deal with the near-extremal case and  we have to investigate how close to the extremal case one can get so that these approximations remain valid. From \eqref{stateequation} it follows that keeping $\left(\frac{M}{\mu}\right)^{2\nu}\ll 1$ and
decreasing 
 the mass $M$, one approaches  closer to  the extremal case.
Our consideration in Appendix A shows that for given parameters $b,c,G,\mu,\nu$
there exists $M_{cr}(b,c,G,\mu,\nu)$ such that for $M>M_{cr}$ all approximations are valid.

\section{Temperature and entanglement entropy with island  under the constraint equation}\label{analysis}

In this section assuming the constraint equation \eqref{stateequation}, we consider the dependence of the temperature of a charged black hole \eqref{temperature} and the entanglement entropy with an island configuration \eqref{islandch} on the mass of the black hole. Using this analysis we study  the behavior of these quantities during  the  evaporation  of  a  charged  black hole under this constraint  \eqref{stateequation}.

\subsection{Asymptotic behavior of the temperature and entanglement entropy with island for small mass}

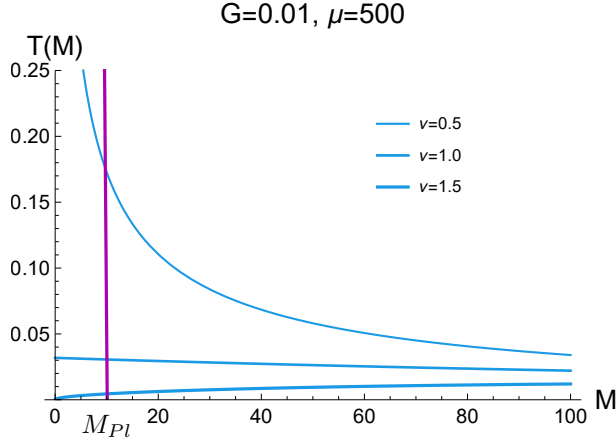
\begin{figure}[h!]\centering
\includegraphics[width=80mm]{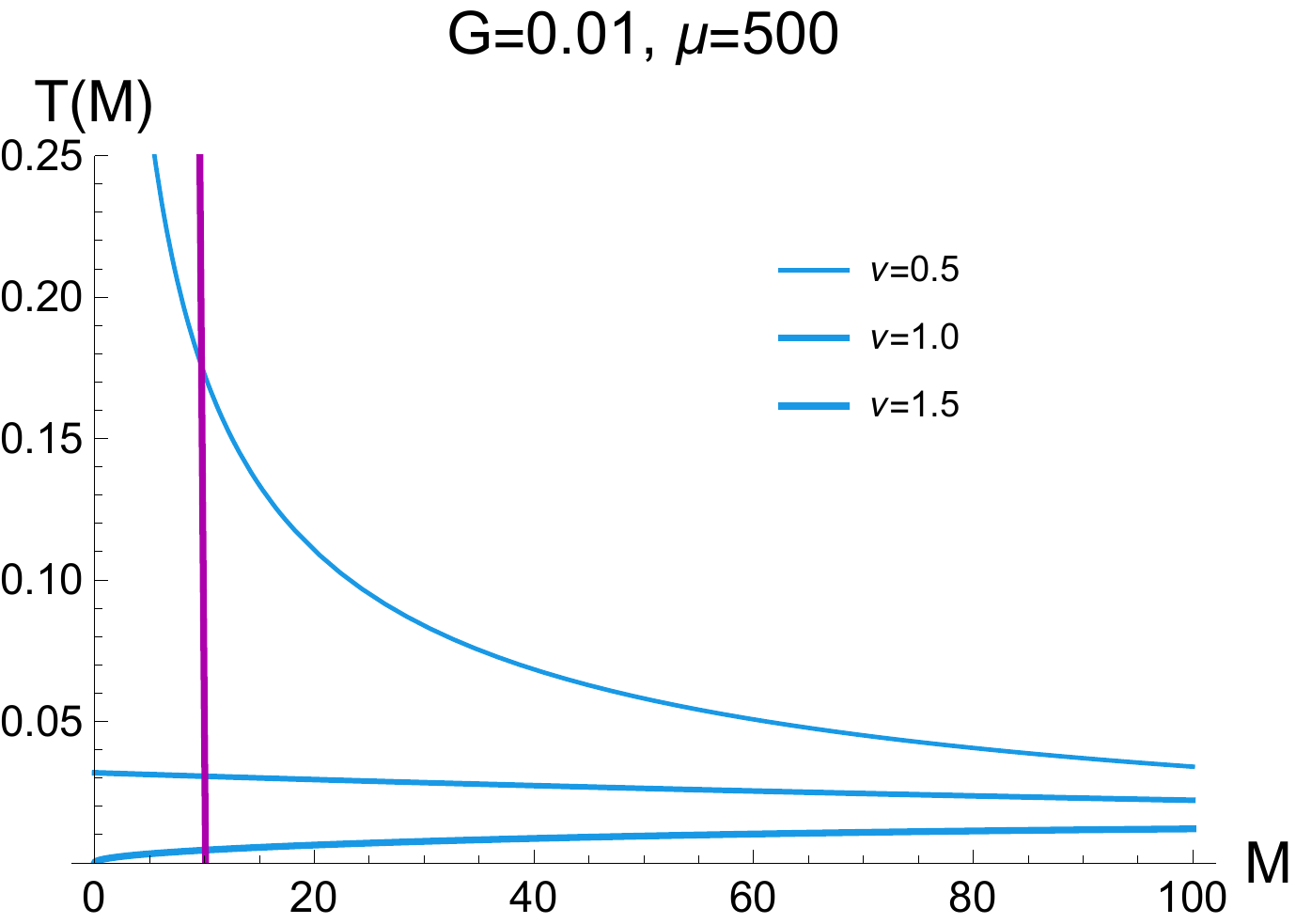}
\begin{picture}(0,0)
\put(-205,-2.0){\footnotesize{$M_{Pl}$}}
\end{picture}
	\caption{The dependence of temperature on mass (blue lines) at $\nu  = 0.5, 1, 1.5$.  For $M \to 0$ the temperature increases to infinity for $\nu<1$, tends to a finite value for $\nu=1$, and tends to zero for $\nu>1$.
	The vertical magenta line corresponds to the Planck mass $M_{Pl} = 1/\sqrt{G}$.}
	\label{temperatura}
\end{figure}

As it can be seen from \eqref{temperature},  the asymptotics  of the temperature at $M\to 0$ depends on the scaling parameter $\nu$. There are the following options for various $\nu$:
\be\label{tempasymp}
\lim _{M\to 0} T=\left\{\begin{array}{cc}
    \infty, & \nu <1,\\
 \displaystyle\frac{1}{2 \pi G \mu}, &  \nu=1,\\
   0, & \nu>1.
\end{array}\right.
\ee
Thus, the temperature $T$  at $M \rightarrow 0$ increases to infinity at $\nu<1$, tends to the finite value at $\nu=1$ and tends to zero at $\nu>1$.  Fig.\ref{temperatura}
 shows temperature vs. mass for various $\nu$.

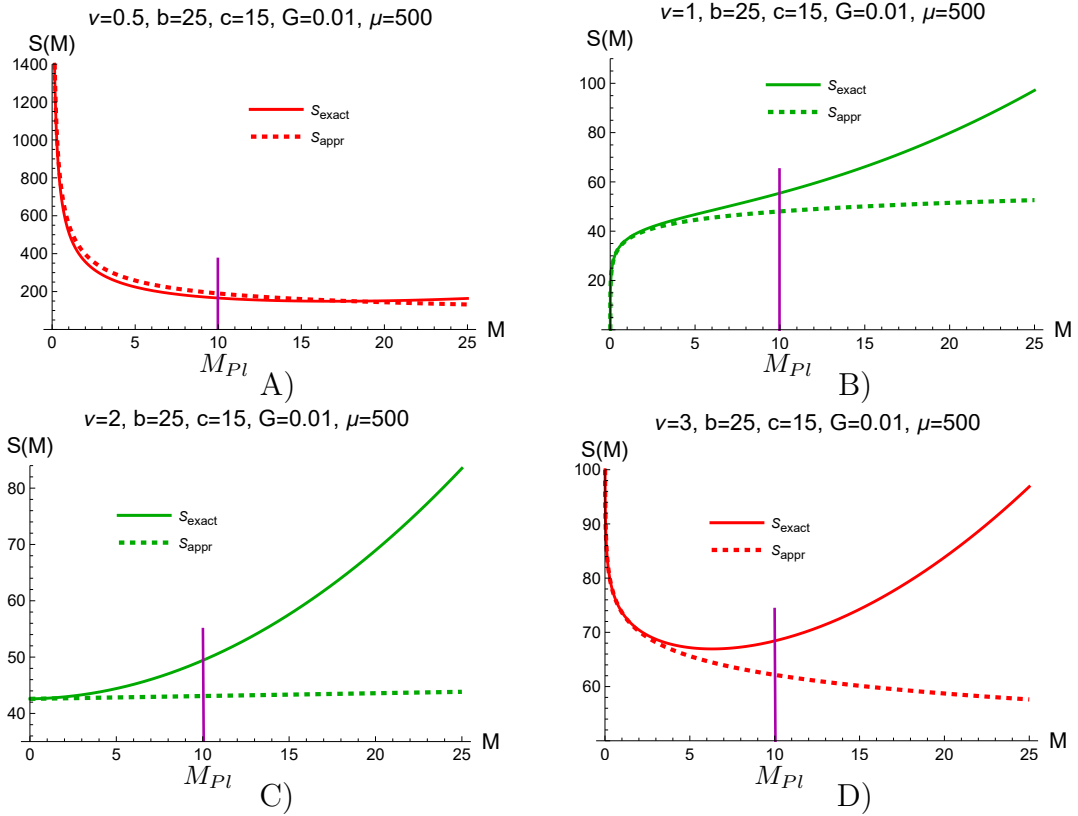
\begin{figure}[h!]\centering
\includegraphics[width=65mm]{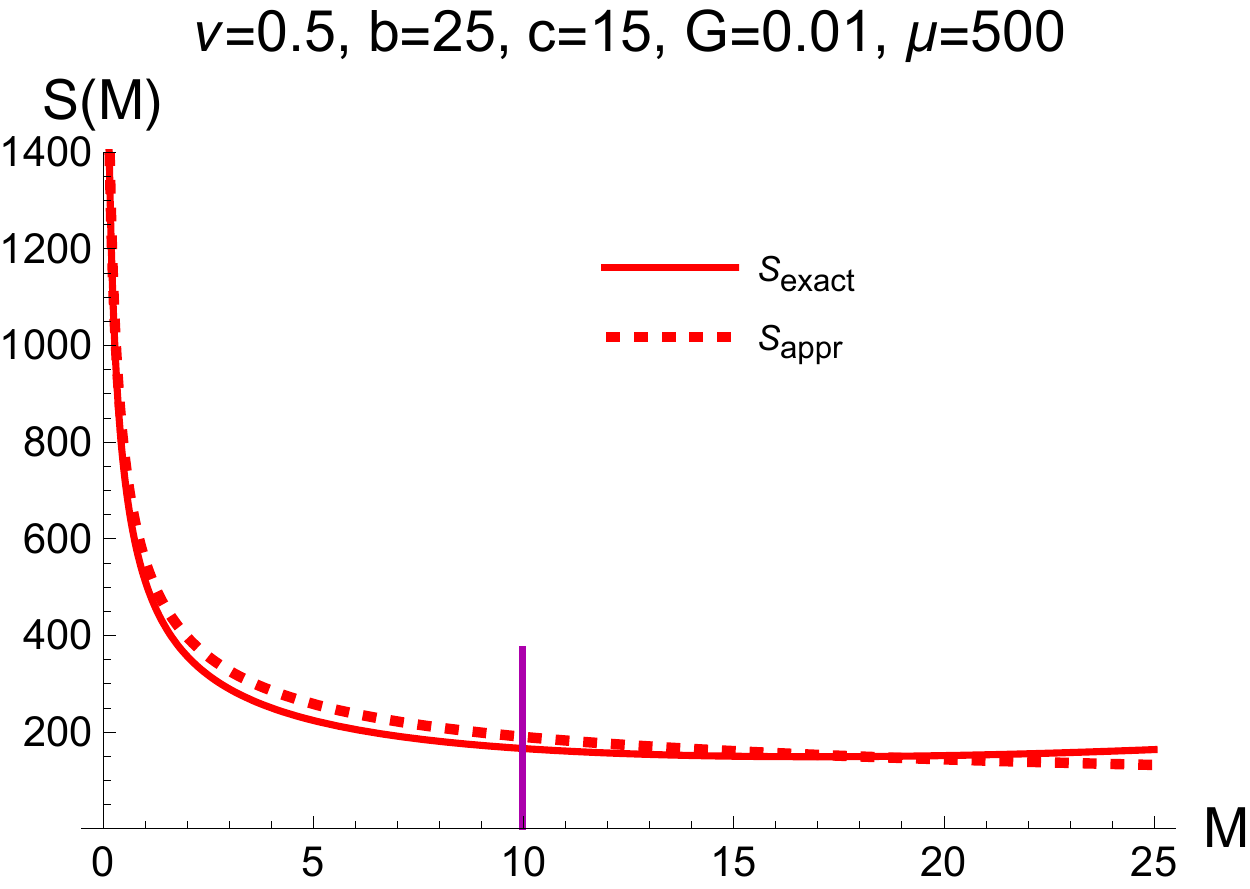}
\qquad\includegraphics[width=65mm]{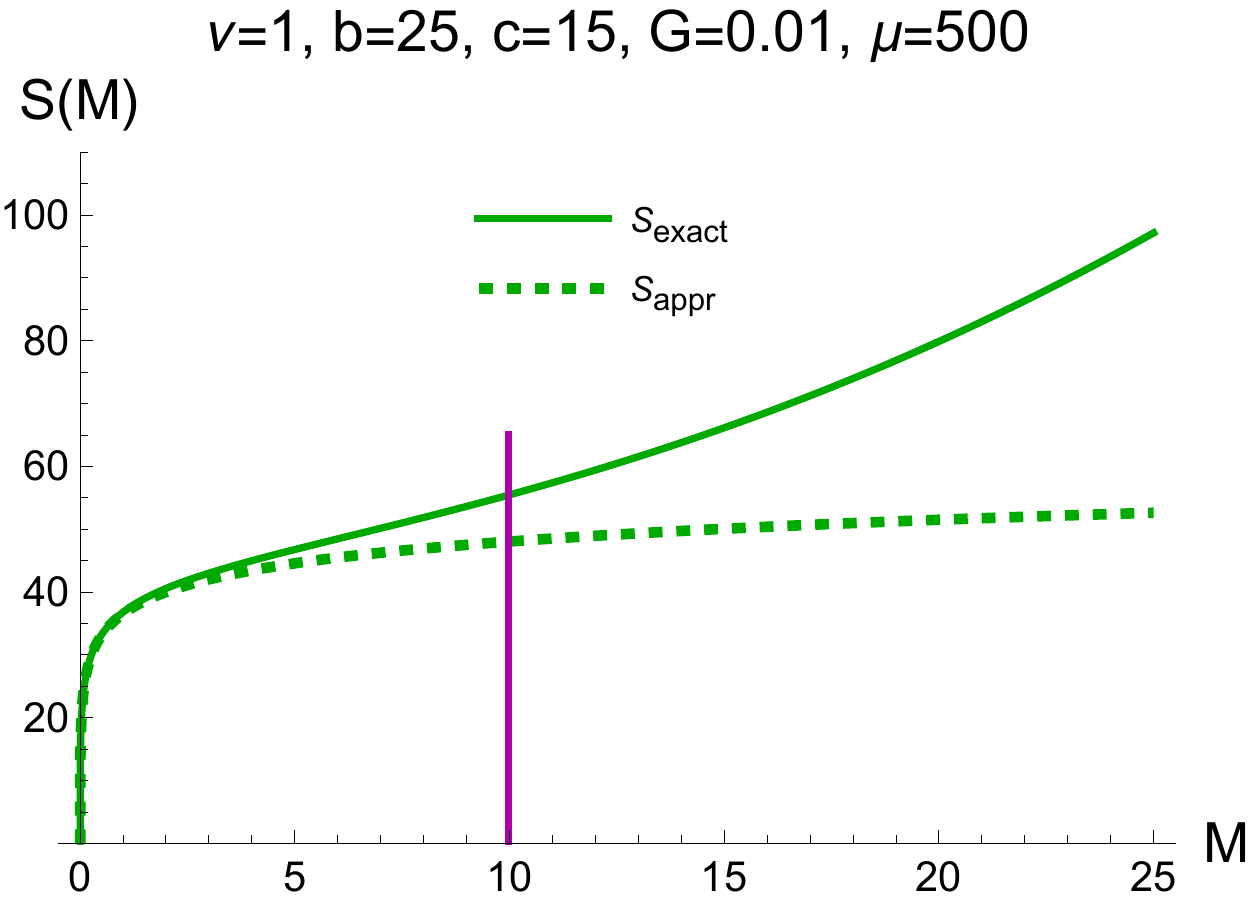}\\ \hspace{5mm} A) \hspace{70mm}B)\\
\includegraphics[width=65mm]{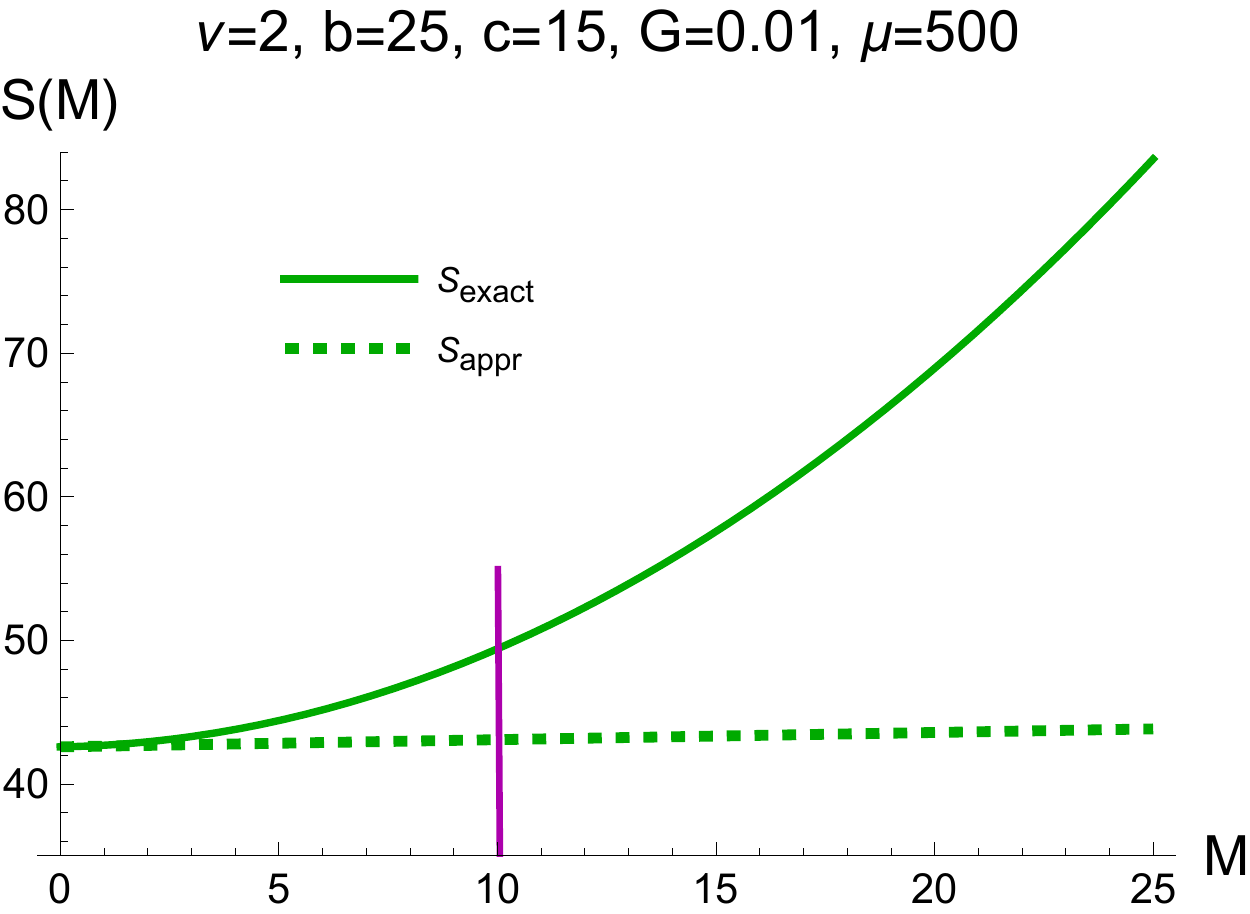}
\qquad\includegraphics[width=65mm]{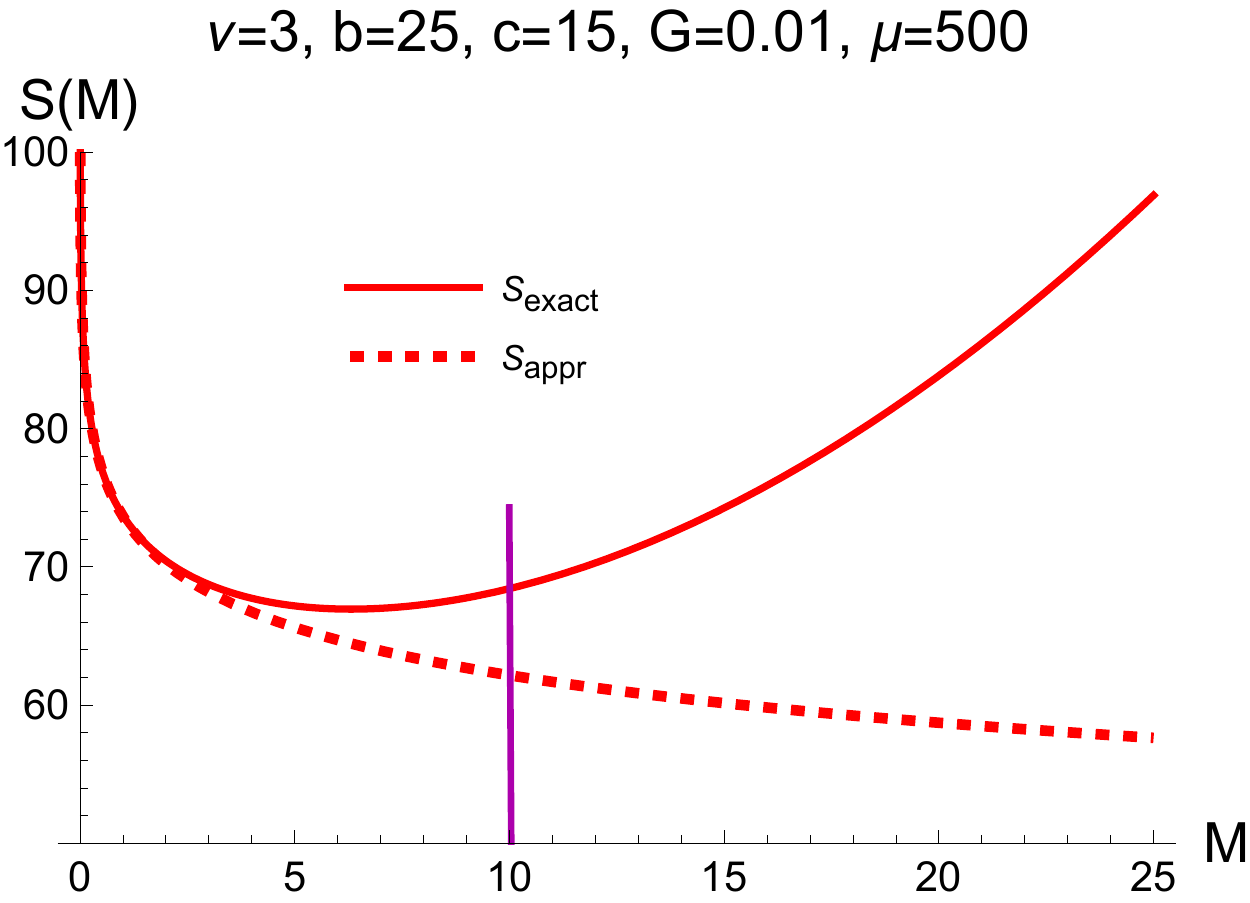}	\begin{picture}(0,0)
\put(-335,-7.5){\footnotesize{$M_{Pl}$}}
\put(-121,-7.5){\footnotesize{$M_{Pl}$}}
\put(-329,148.0){\footnotesize{$M_{Pl}$}}
\put(-118,148.0){\footnotesize{$M_{Pl}$}}
\end{picture}\\ \hspace{5mm} C) \hspace{70mm}D)\\
	\caption{Mass dependence
	 of exact \eqref{entropwithisland} and approximate (\ref{asymptorics}) entropy with the island  at $\nu=0.5, 1, 2, 3$.
	 Depending on the presence or absence of an infinite increase at $M \to 0$, see \eqref{entropyasymp}, the plots are shown in red A), D) or green B), C), respectively.
	 The case C) with $\nu=2$ is unique because the entropy with the island is finite in this case for $M=0$.
 The vertical magenta line corresponds to the Planck mass $M_{Pl} = 1/\sqrt{G}$. }%{\bf Math.file: Derivative-Verification-TR-16-12.nb}}
	\label{smallM}
\end{figure}

\begin{figure}[h!]\centering
\includegraphics[width=180mm]{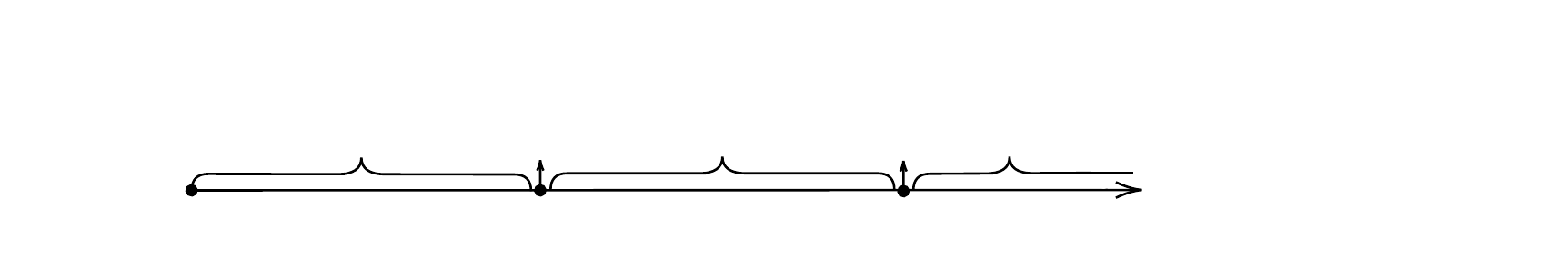}
\begin{picture}(0,0)
\put(-59,89){\large{Asymptotics at $M \to 0$}}
\put(-156,28.5){$\Huge{0}$}
\put(-42,28.5){$\Huge{1}$}
\put(76.9,28.5){$\Huge{2}$}
\put(149,28.95){\large{$\nu$}}
\put(4.5,66){\small{$T \to 0$}}
\put(0,56){\small{$S_{\cI}  \to -\infty$}}
\put(-112.5,66){\small{$T \to \infty$}}
\put(-117,56){\small{$S_{\cI}  \to \infty$}}
\put(121,66){\small{$T \to 0$}}
\put(117,56){\small{$S_{\cI}  \to \infty$}}
\put(-58.5,66){\small{$T = \text{finite}$}}
\put(-60.5,56){\small{$S_{\cI}  \to -\infty$}}
\put(62,66){\small{$T \to 0 $}}
\put(56,56){\small{$S_{\cI}  = \text{finite}$}}
\end{picture}
	\caption{Asymptotic behavior at $M\to 0$ of the black hole temperature \eqref{tempasymp} and  entanglement entropy with island \eqref{entropyasymp} for various $\nu$. There are 5 different regions: [0,1), 1, (1,2), 2, $(2,+\infty)$. The segment $[1,2]$ is the most preferable from the point of view of the absence of growth at $M\to 0$.
	}
	\label{comparison}
\end{figure}
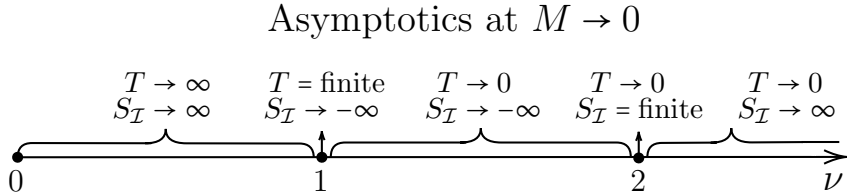
$$\,$$

Assuming $\alpha\ll 1$ and $G M\ll b$ we get in the leading order at $M\to 0$ for the entropy with the island \eqref{entropwithisland}

   \bea\label{asymptorics}
S^{leading}_{\cI,appr} &=& \frac{c
	b M^{\nu-1}}{3G \mu^{\nu}}+ \frac{c}{6} \log \left(4 G^2 M^{4-2\nu} \mu^{2 \nu}\right).
\eea
It can be seen from (\ref{asymptorics}) that the asymptotic behavior of the entropy with the island for $M\to 0$ depends on the scaling parameter $\nu$. For different~$\nu$ there are the following options:
\bea\label{entropyasymp}
\lim _{M\to 0} S^{leading}_{\cI,appr} =
\left\{\begin{array}{cc}
    \infty, &  0\leq \nu <1,\\
    -\infty, &  1\leq\nu<2,\\
   \displaystyle\frac{c}{3} \log \left( 
2 G\mu ^2\right), &  \nu=2,\\
   \infty, & \nu>2.
\end{array}\right.
\eea
Thus, at $M\to 0$ the entropy with island (\ref{entropy}) at $\nu\in[0,1) \cap (2, +\infty)$ grows to infinity, at $\nu \in [1,2)$ it decreases to minus infinity and at $\nu=2$ it takes the finite value.
Fig.\ref{smallM} shows this behavior of entropy with the island for relatively small masses.  
\\

A comparison of asymptotics of the temperature and asymptotics of the entropy with the island at $M\to 0$ is shown in Fig.\ref{comparison}.
\\

\subsection{Entropy derivative}

 In order to investigate the issue of the occurrence of entropy growth, including the region near the Planck mass $M > M_{Pl}$, it is useful to consider  the sign of the mass derivative of entropy with island $\partial _MS_{\cI}$. It is clear that during the black hole evaporation one has
\bea
\partial_t S_{\cI} &=& \partial_M S_{\cI} \frac{dM}{dt}, \\
\frac{dM}{dt} &<& 0, \; \; t>0,
\eea
therefore $\partial_M S_{\cI}>0$ ($\partial_M S_{\cI}<0$) corresponds to a decrease (increase) of entropy with time. 
\\

\begin{figure}[h!]\centering
\includegraphics[width=70mm]{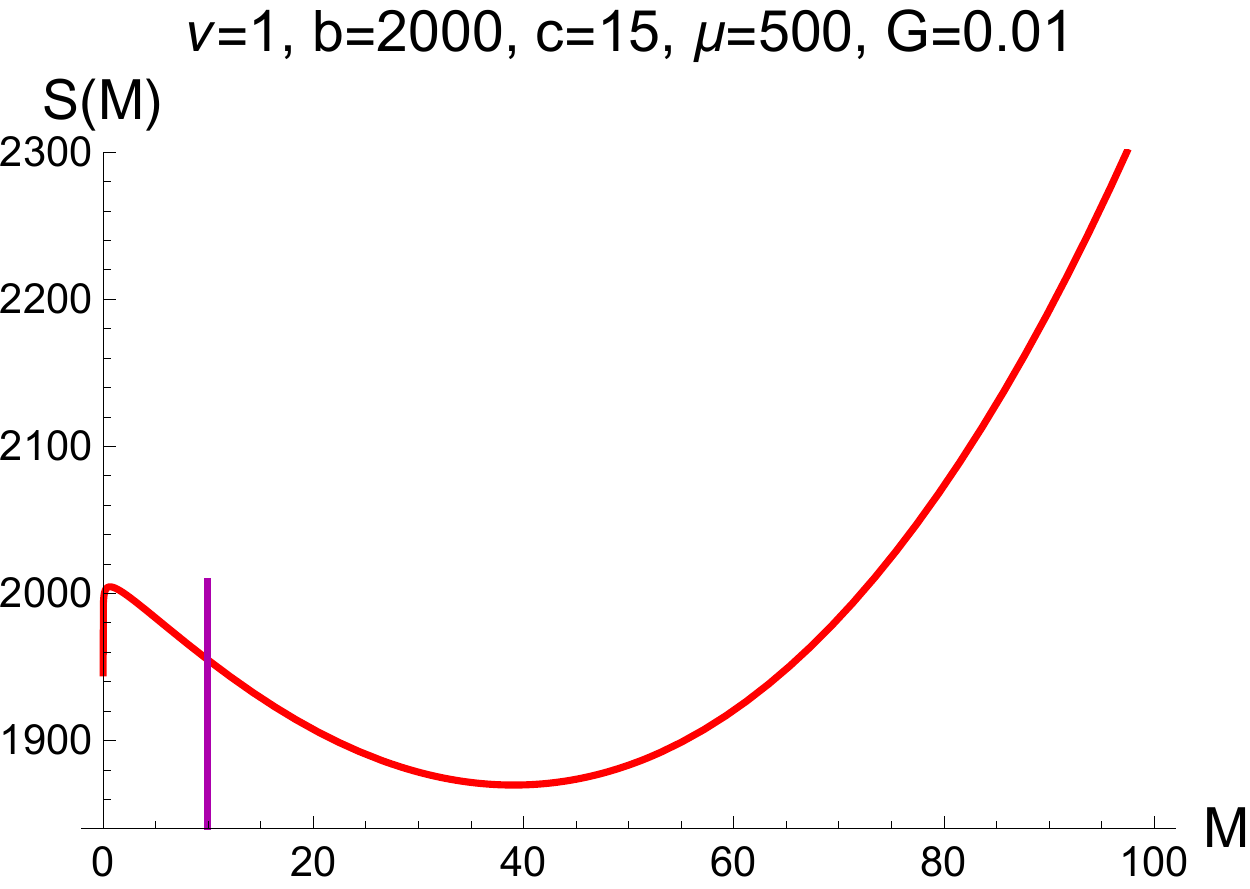}
	\qquad\includegraphics[width=70mm]{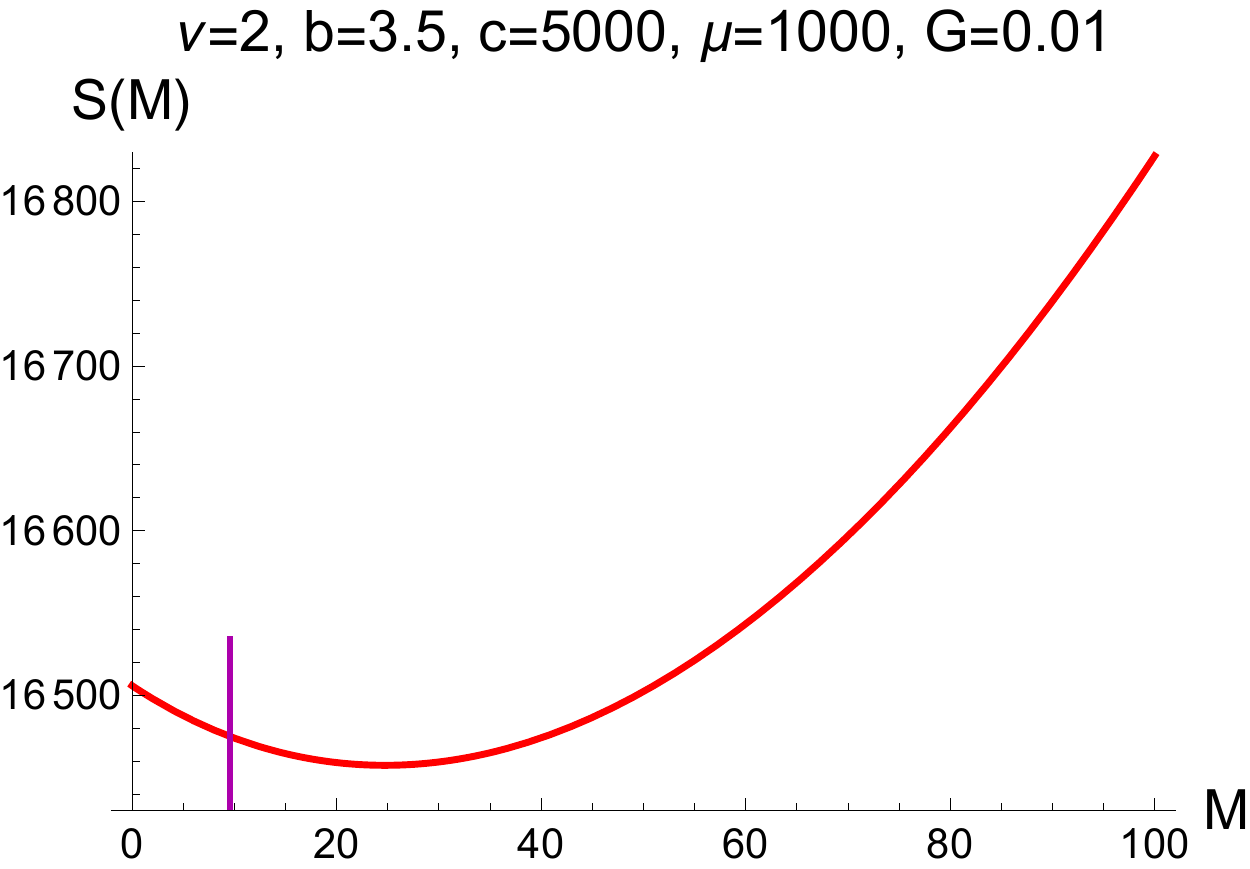}\begin{picture}(0,0)
\put(-400,-1.5){\footnotesize{$M_{Pl}$}}
\put(-171,-1.5){\footnotesize{$M_{Pl}$}}
\end{picture}\\ \hspace{-10mm} A) \hspace{75mm}B) \\
	\caption{The dependence of entropy with island on mass (red lines) at $\nu=1$ and $\nu=2$ (A) and B), respectively) with such ratios of the parameters $b$, $\mu$ and $c$, that there is the increase of entropy
	with decreasing of mass in the region $M>M_{Pl} = 1/\sqrt{G}$  above the Planck mass (the vertical magenta line).}% {\bf Math.file: Derivative-Verification-TR-16-12.nb}}
	\label{nu12growth}
\end{figure}
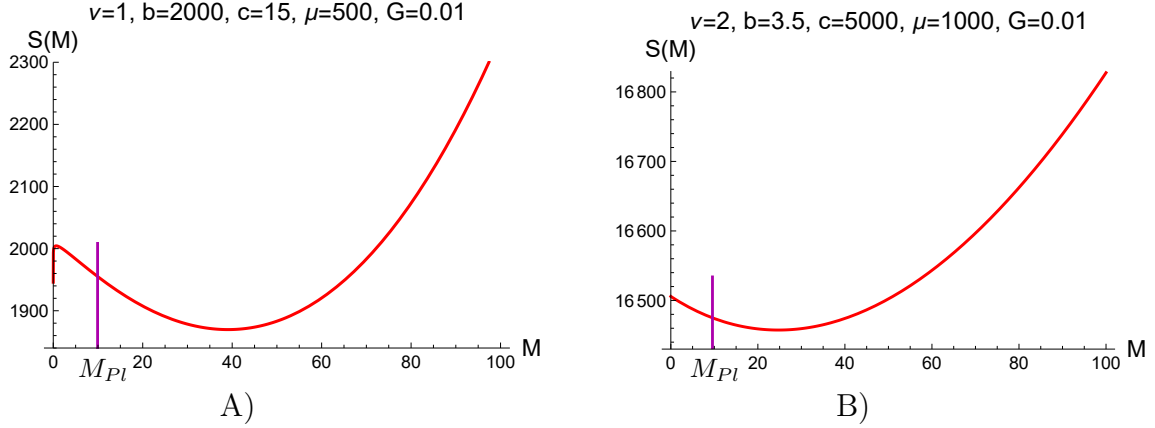

The mass derivative of entropy with the island (\ref{entropy}), multiplied by the mass $M>0$ and 
expanded over small $\alpha$ and $\tau\equiv \frac{GM}{b}$ is
\bea\label{derivative}
M \partial_M S_{\cI
} &=& 4 \pi  G M^2
  + \frac{c \alpha}{3 \tau} \left(\nu-1\right)+\frac{2 c \alpha^2  (1-2\nu) }{3 \tau} + \frac{c \left(2-\nu\right)}{3}
-\frac{c \tau}{3}\\
&+&\frac{2 c \nu \alpha}{3}  \log\left(  \frac{1}{2 \alpha \tau} \right)+ \cO(\alpha)+\cO(\tau^2)+\cO\left(\frac{\alpha^3}{\tau}\right)+\cO(\alpha^2 \log \alpha \tau). \nn
\eea
Let us consider in more details  the segment $\nu \in [1,2]$. It can be seen that in (\ref{derivative}) in this case only the third term of order $\cO \left(\displaystyle\frac{\alpha^2}{\tau}\right)$ and the fifth term of order $\cO(\tau)$ are negative.
These terms can lead to negative derivative \eqref{derivative} in some right semi-neighbourhood of $\nu=1$ and left semi-neighbourhood of $\nu =2$ for certain ratios of the parameters $\alpha$, $\tau$ and $c$.\footnote{Indeed, at $\nu \in [1,\cV_1)$ for some $\cV_1$ when $\tau$ decreases (if $b$ increases), increasing negative third term of order $\cO \left(\displaystyle\frac{\alpha^2}{\tau}\right)$  can lead to negative derivative \eqref{derivative}. Similarly, at $\nu\in (\cV_2,2]$ for some $\cV_2$ when $\alpha$ decreases (if $\mu$ increases) and $c$ is large enough to exceed the first term, negative fifth term of order $\cO(\tau)$ can lead to negative derivative \eqref{derivative}.} Thus, despite that the entropy $S_{\cI}$ does not increase at $\nu \in [1,2)$ at $M \to 0$ and at $\nu=2$ is finite at $M=0$, see \eqref{entropyasymp}, in the region of masses above the Planck mass $M > M_{Pl}$, entropy growth is possible.
Fig.\ref{nu12growth} demonstrates entropy growth at $\nu=1$ and $\nu=2$ for specially chosen ratios of the parameters $b$, $\mu$ and $c$.

Note that   we can make a rough estimate for the domain $\nu \in [1,2]$ where the derivative~(\ref{derivative}) is non-negative for some given values of the parameters $b$, $\mu$ and $c$.
The non-negativity of the sum of the second term of order $\cO \left(\displaystyle\frac{\alpha}{\tau}\right)$ and the third term of order $\cO \left(\displaystyle\frac{\alpha^2}{\tau}\right)$ from (\ref{derivative}) gives the estimate:
\be\label{firstbound}
\nu\geq\frac{1-2 \alpha}{1-4  \alpha} =\frac{1-2 \left(\frac{M}{\mu} \right)^{\nu}}{1-4 \left(\frac{M}{\mu} \right)^{\nu}} \simeq 1+2 \left(\frac{M}{\mu} \right)^{\nu}.
\ee

The non-negativity of the sum of the fourth term of order $\cO (1)$ and the fifth term of order $\cO (\tau)$ from (\ref{derivative}) gives the estimate:
\be\label{secondbound}
\nu\leq2-\tau=2-\frac{G M}{b}.
\ee
Let us take in inequalities (\ref{firstbound}) and (\ref{secondbound}) maximum value of the mass $M_{\text{max}}$ and denote the values of $\nu$ at which these inequalities are saturated as $\nu_1$ and $\nu_2$, respectively. We get the transcendental equation for $\nu_1$ and equation for $\nu_2$
\bea\label{firstbound-tr}
\nu_1 &=&1+2 \left(\frac{M_{\text{max}}}{\mu} \right)^{\nu_1},
\\
\nu_2&=&2-\frac{G M_{\text{max}}}{b}.
\eea

If we consider that the maximum mass is taken and that in the derivative (\ref{derivative}) there are other unaccounted positive terms, then approximately $\nu_1$ is an upper estimate and $\nu_2$ is an lower estimate for the boundaries of the segment in which the derivative is non-negative. Thus, the segment $[\nu_1, \nu_2]$ is embedded in the actual segment $[\cV_1, \cV_2] \subset [1,2]$ where $\partial_M S_{\cI} \geq 0$.
\\

\begin{figure}[h]\centering
\includegraphics[width=180mm]{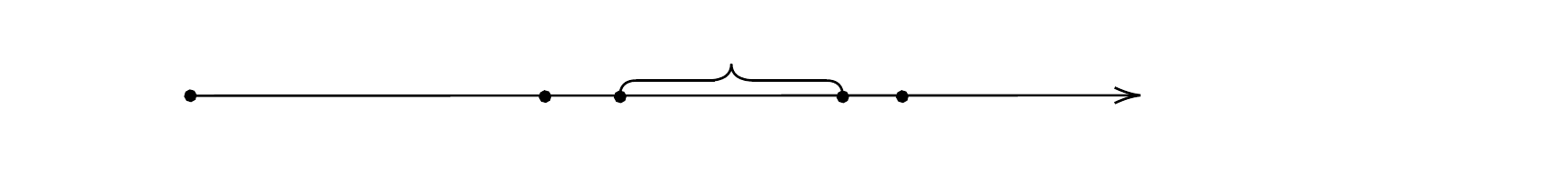}
\begin{picture}(0,0)
\put(-156,27){$\Huge{0}$}
\put(-39.9,27){$\Huge{1}$}
\put(-17,28.0){$\Huge{\cV_1}$}
\put(55,28){$\Huge{\cV_2}$}
\put(76.9,27){$\Huge{2}$}
\put(149,28){\large{$\nu$}}
\put(-2,56){$\Huge{\partial_M S_{\cI}\geq 0}$}
\end{picture}
	\caption{Segment $[\cV_1, \cV_2] \subset [1,2]$ on the $\nu$-axis, for which the mass derivative of entropy with an island $\partial_M S_{\cI}\geq 0$ is non-negative, exists for any physically meaningful values of the parameters.}
	\label{nuaxis}
\end{figure}
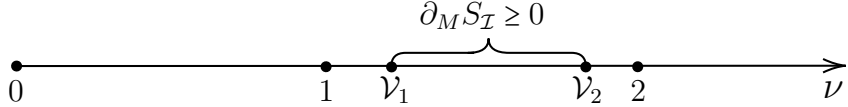

The values $\nu_1$ and $\nu_2$ depend on the parameters $\mu$ and $b$, respectively. However, due to the smallness of $\left(\frac{M}{\mu} \right)^{\nu} \ll 1$ and $\frac{GM}{b} \ll 1$ nontrivial region $[\nu_1,\nu_2]$ exists for any physically meaningful values of $b$, $\mu$ and $c$ and for the entire mass range, i.e. $\nu_1 < \nu_2$ is always satisfied. The same is true for the segment $[\cV_1, \cV_2] \subset [1,2]$.  This is illustrated by Fig.\ref{nuaxis}.
\\

It can be seen from \eqref{derivative} that at $\nu<\cV_1$ for sufficiently small $\tau$ (large $b$), and at $\nu>\cV_2$ for sufficiently small $\alpha$ (large $\mu$ and/or $\nu$) and large $c$, the mass derivative of entropy with island is negative $\partial_M S_{\cI} < 0$. However, as shown in Appendix \ref{appendixapproximation}, for very small $\alpha$ and large $c$, the approximations under which the entropy with an island \eqref{islandch} is derived become invalid. Therefore, it seems that the increase in entropy with an island at $\nu>\cV_2$ occurs at masses $M < M_{cr}$ less than the critical mass.

\begin{figure}[h!]\centering
	\includegraphics[width=42mm]{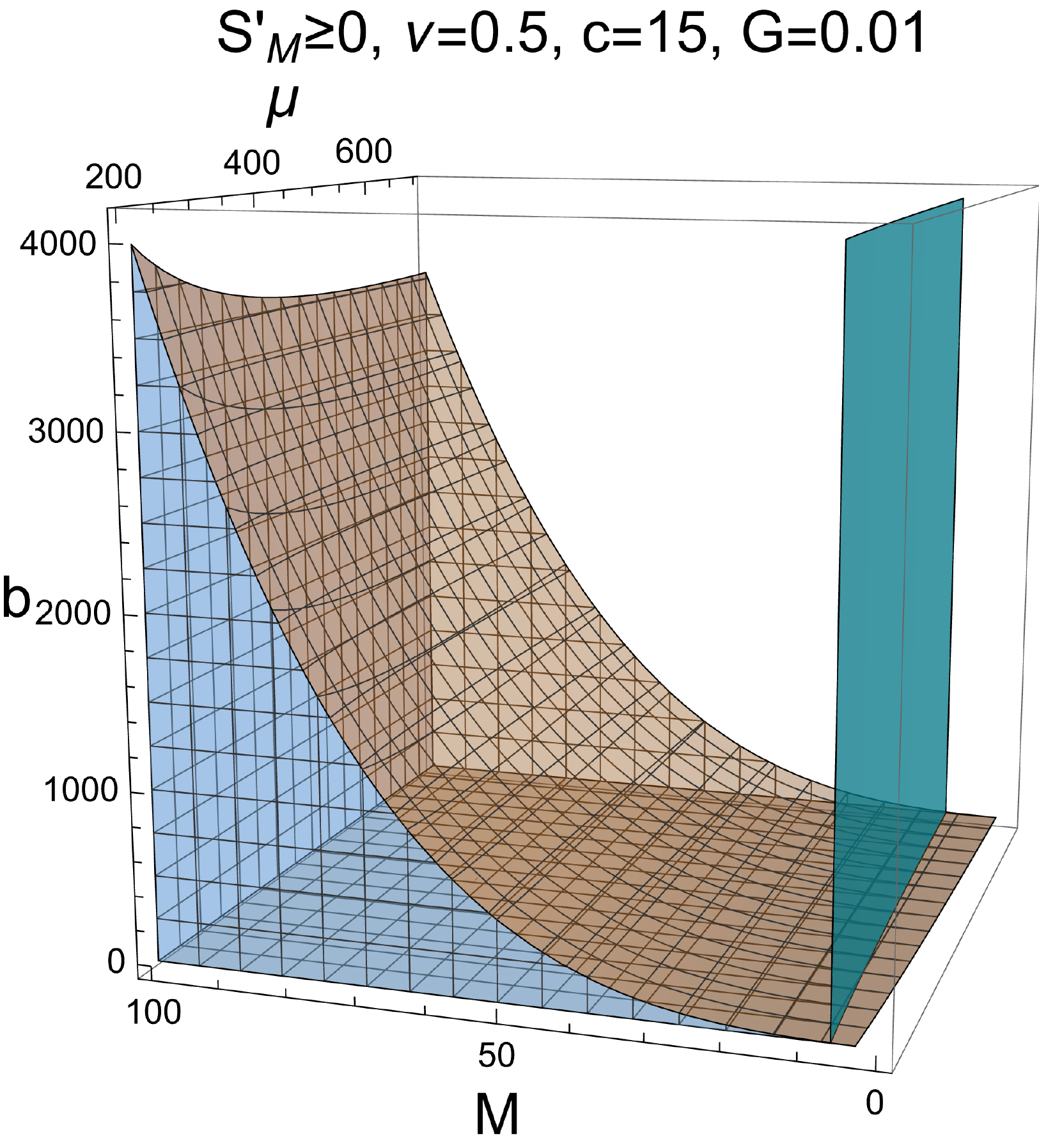}\,A)\qquad\qquad
	\qquad\includegraphics[width=62mm]{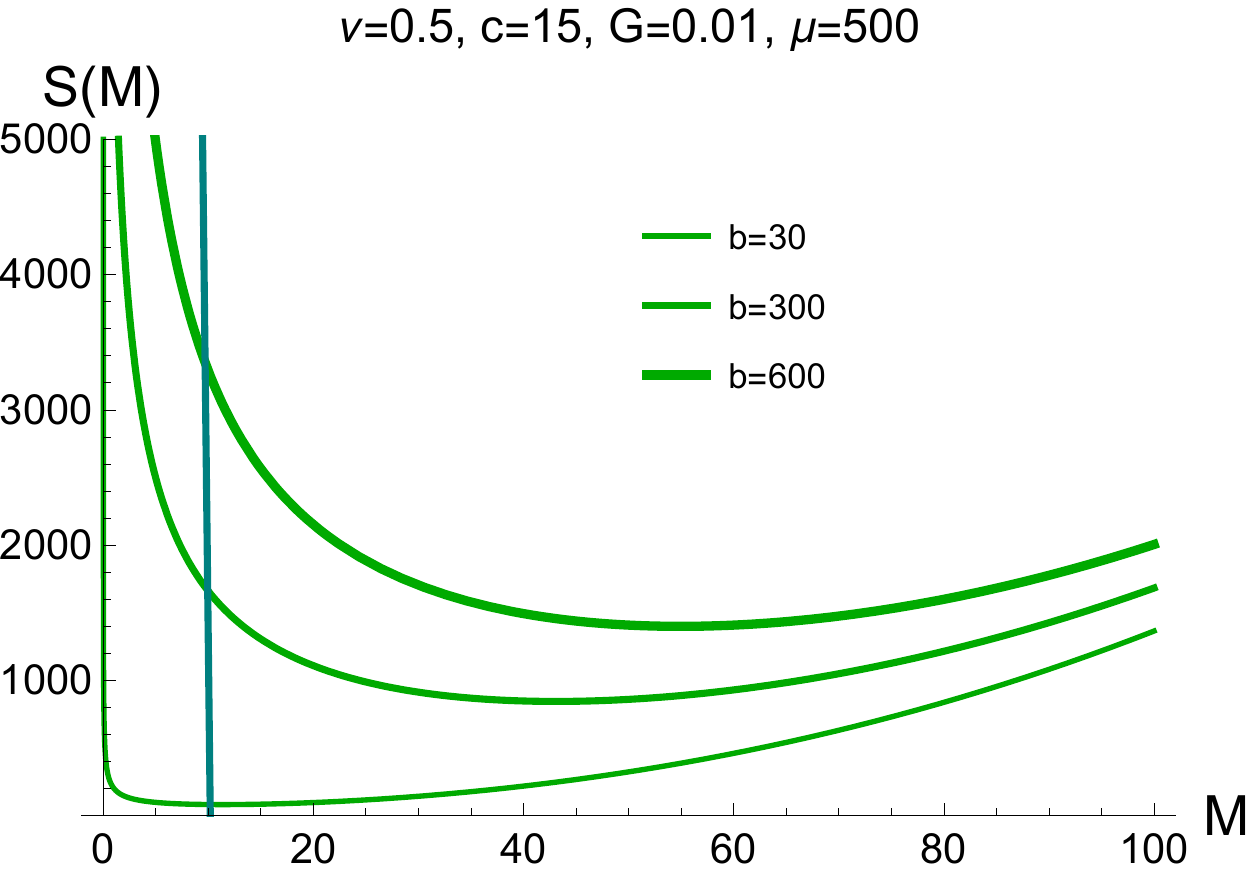}\,B)\\
	\includegraphics[width=42mm]{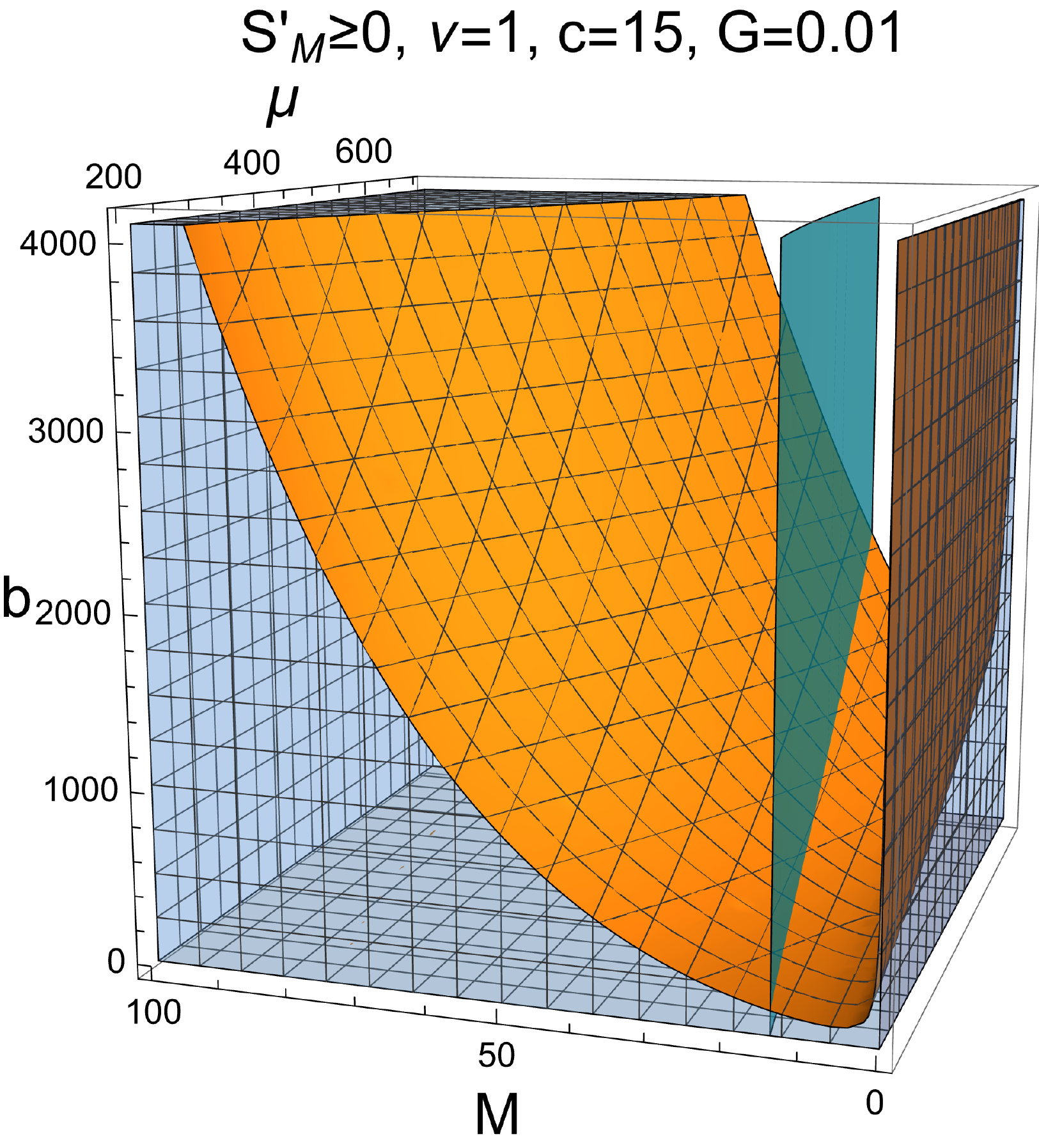}\,C)\qquad\qquad
	\qquad\includegraphics[width=62mm]{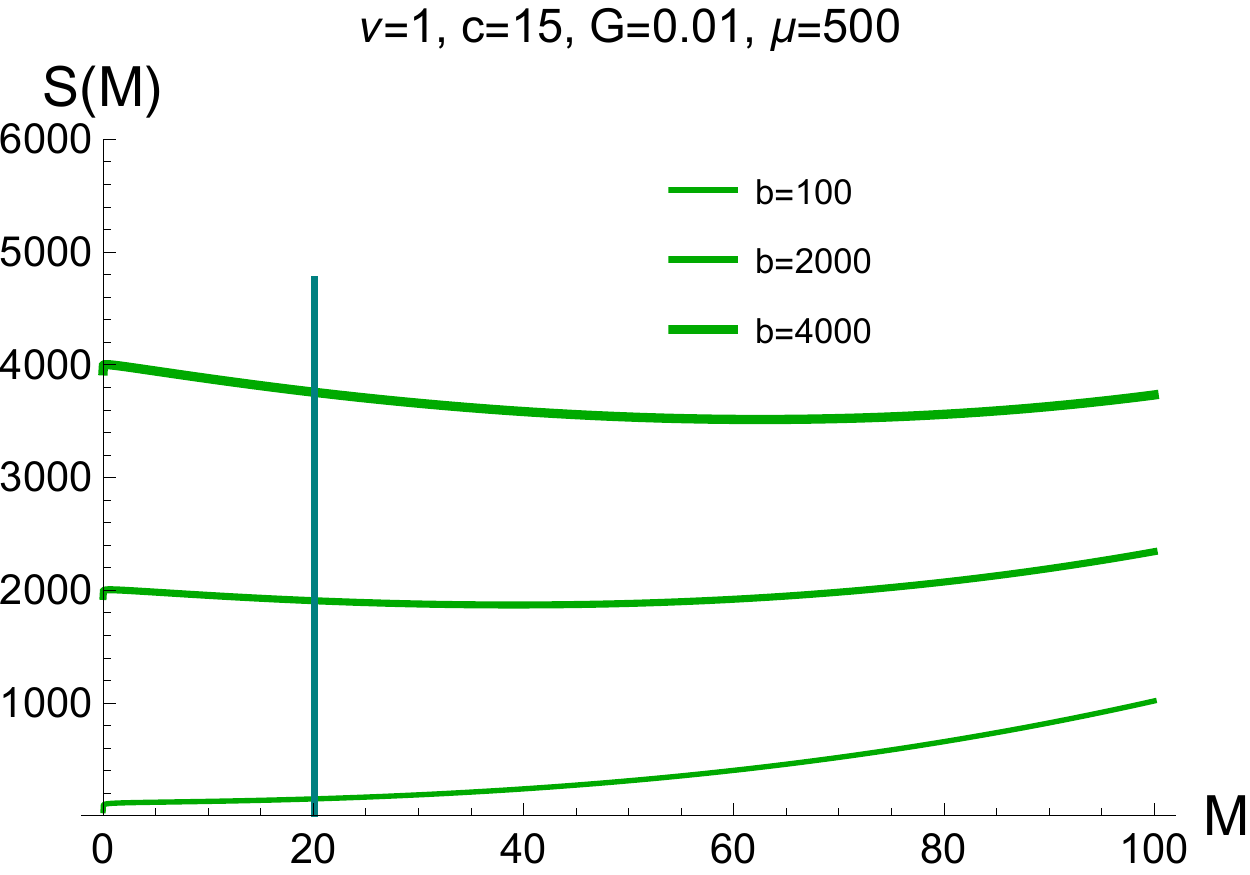}\,D)\\
	\includegraphics[width=42mm]{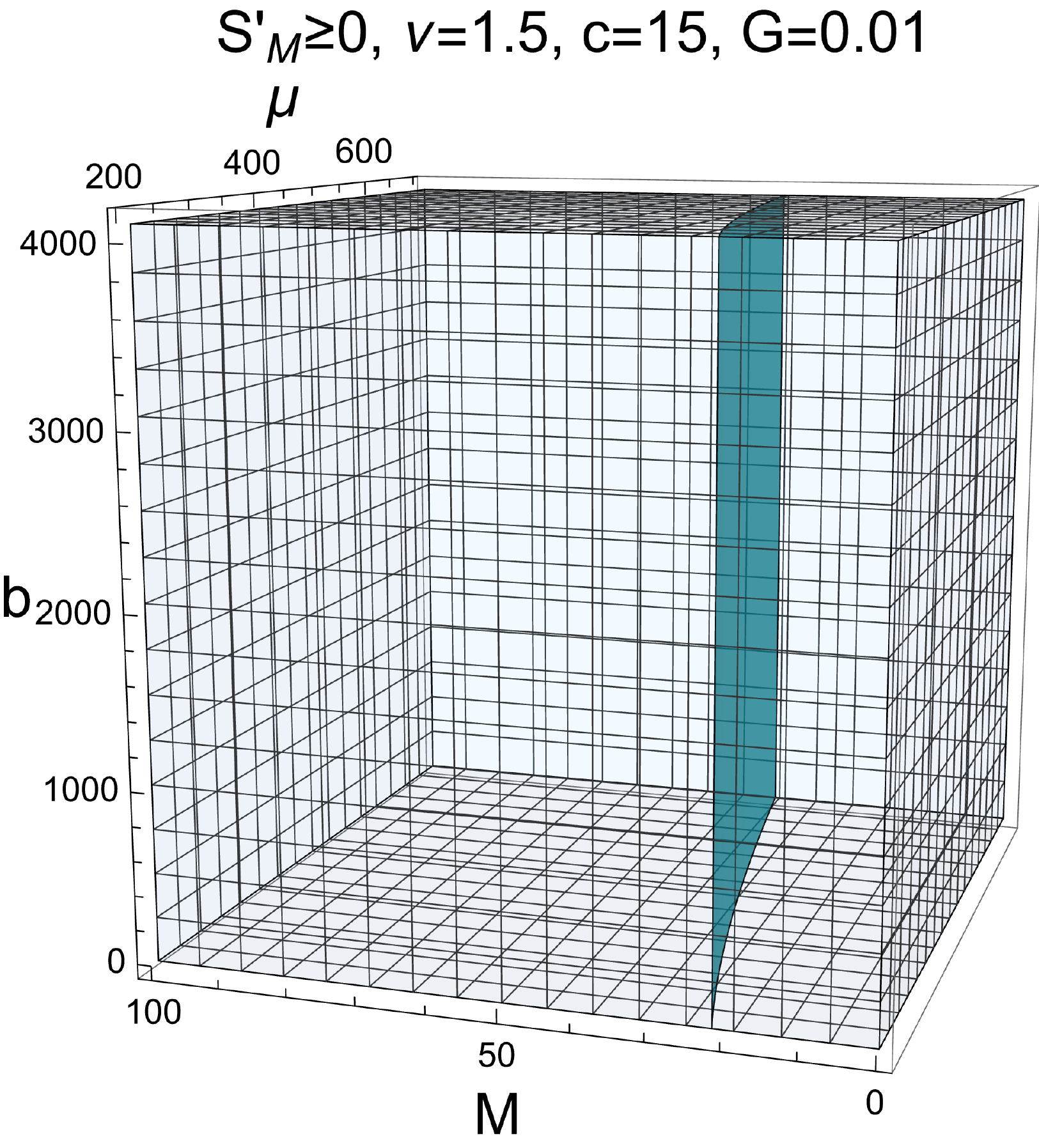}\,E)\qquad\qquad
	\qquad\includegraphics[width=62mm]{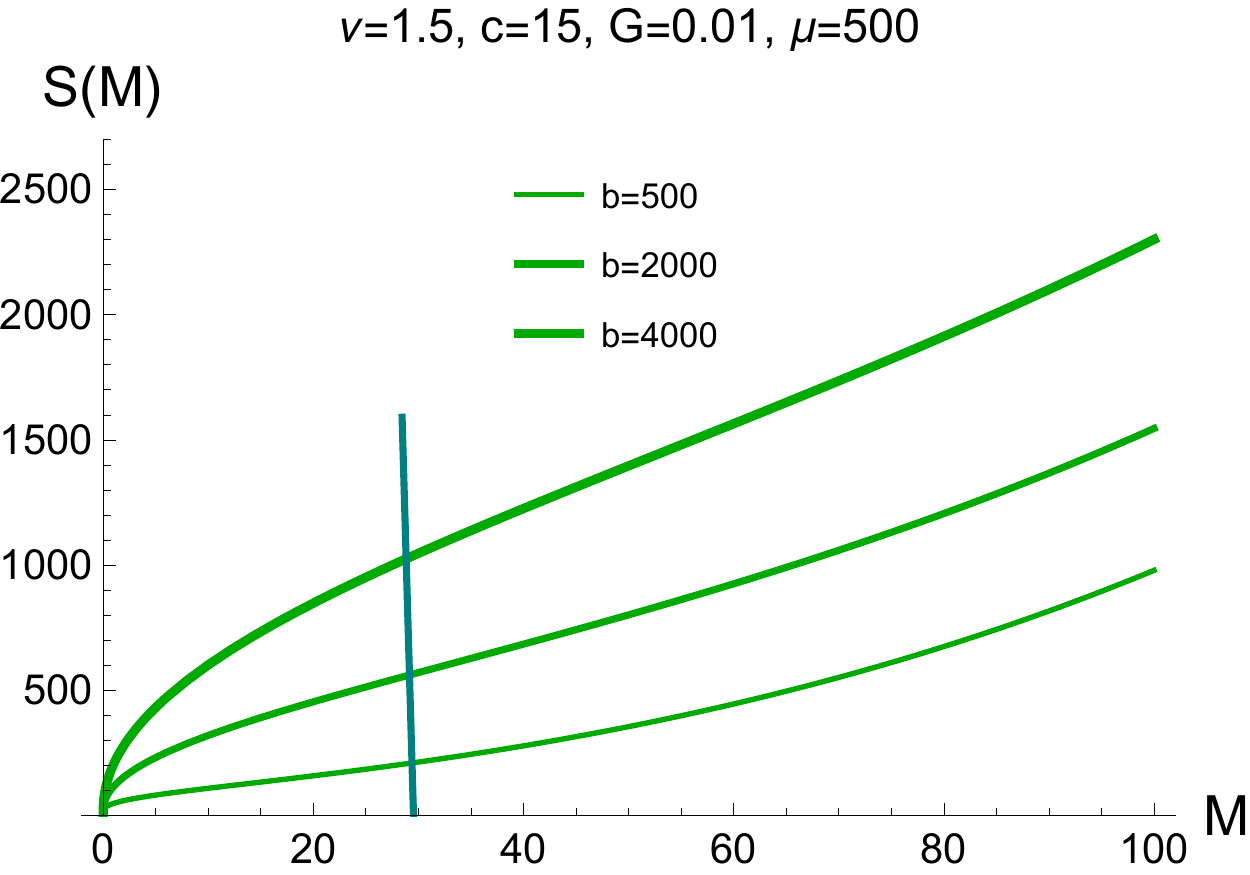}\,F)\\
	\includegraphics[width=42mm]{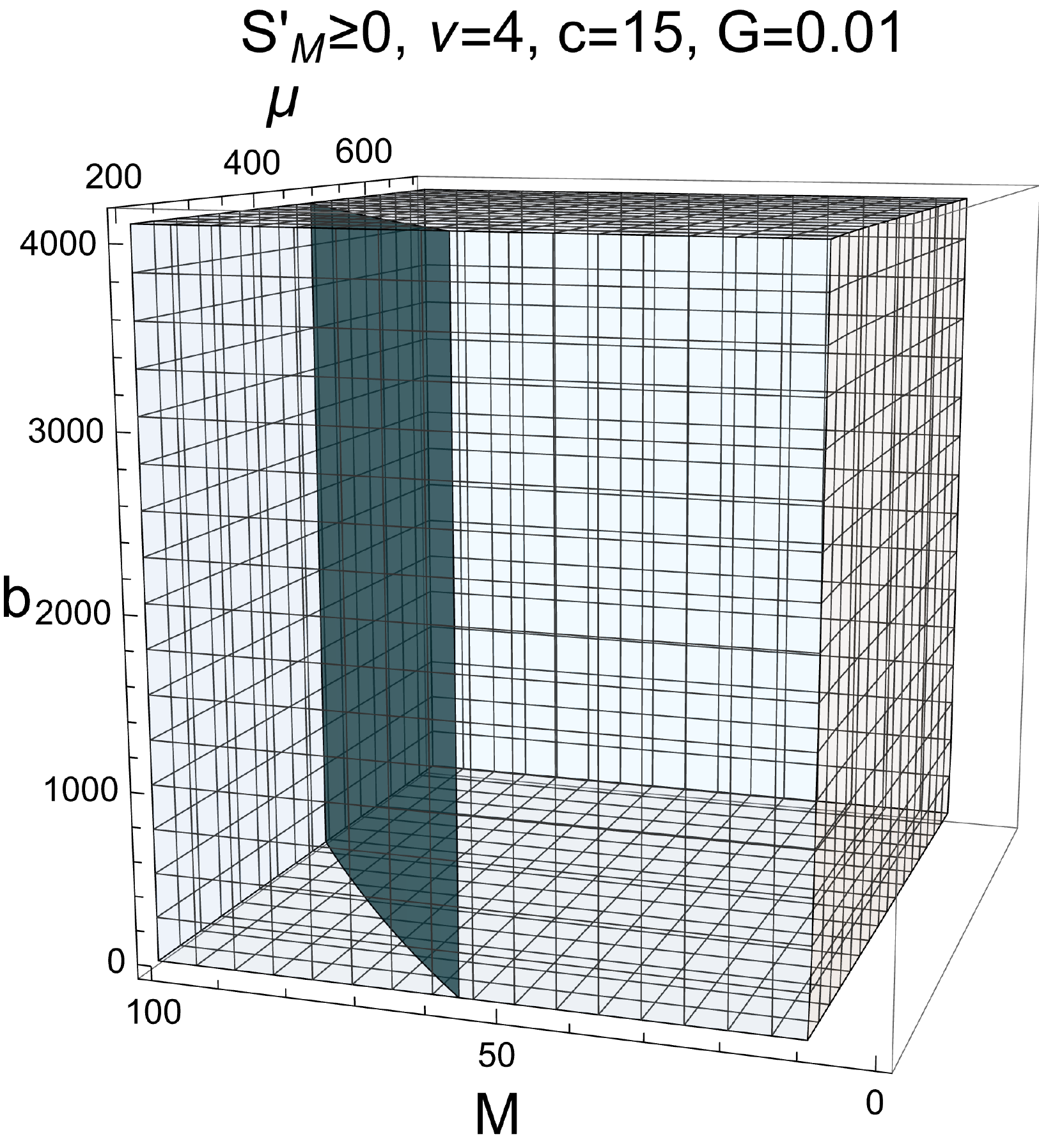}\,G)\qquad\qquad
	\qquad\includegraphics[width=62mm]{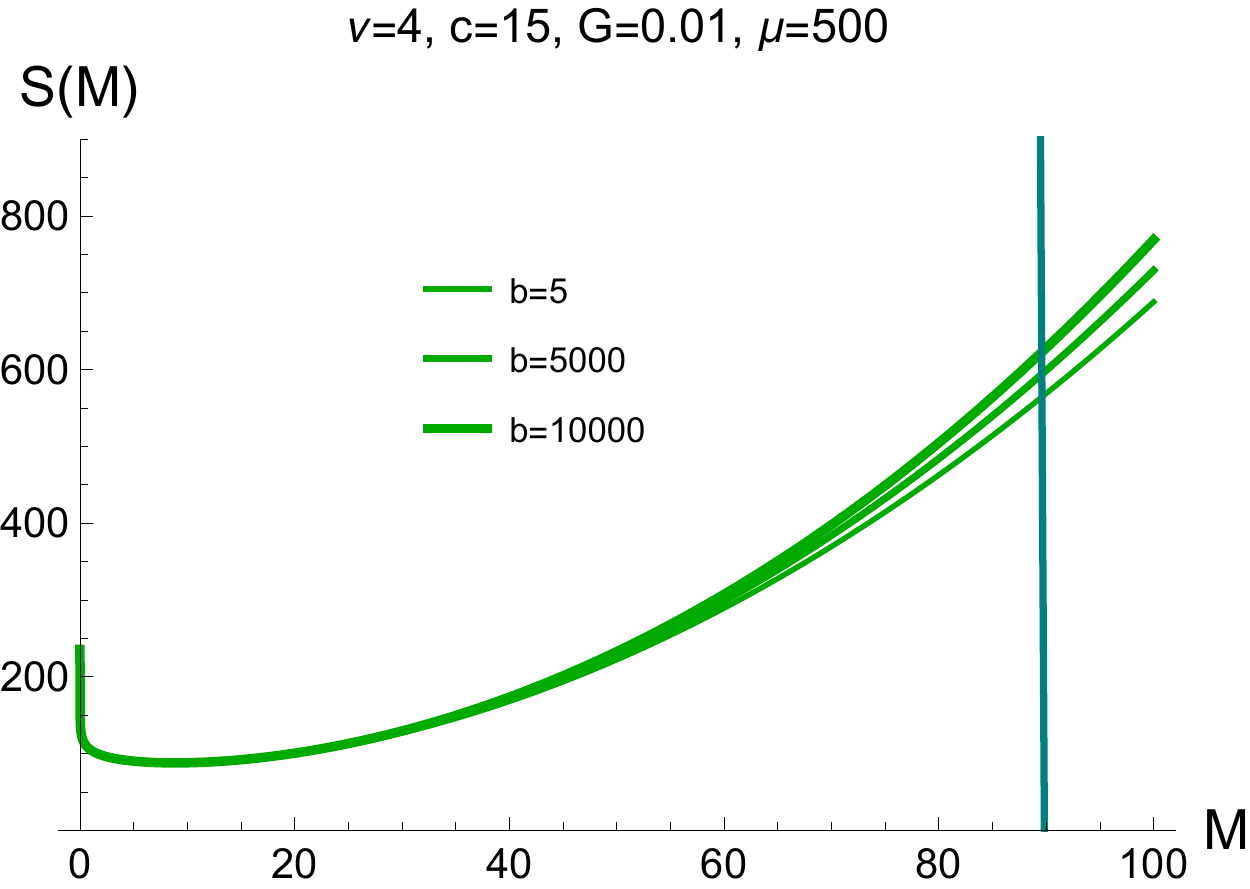}\,H)\\
	\caption{The left column corresponds to numerical analysis of the sign of the mass derivative of entropy with an island $\partial_M S_{\cI
}$ at different $\nu=0.5, 1, 1.5,4$ depending on $M$, $b$ and $\mu$. The right column corresponds to the entropy with the island $S_{\cI}$ depending on mass $M$ of the black hole at the same  $\nu$ and different $b$. The cyan surfaces corresponds to the critical mass $M>M_{cr}$,  up to which the evaporation of a black hole can be considered.}
	\label{threedimensionalgraph}
\end{figure}
$\,$\\

The plots in the left column in Fig.\ref{threedimensionalgraph} show the mass derivative of entropy with an island $\partial_M S_{\cI
}$ at various $\nu=0.5, 1, 1.5,4$ depending on the $M$, $b$ and $\mu$:  the blue shaded areas correspond to the regions of non-negativity of the derivative $\partial_M S_{\cI
} \geq 0$, which implies no increase in entropy with an island over time. The case $\nu=1.5$ which almost certainly belongs to $[\cV_1, \cV_2]$ corresponds to a positive derivative for all values of the parameters (Fig.\ref{threedimensionalgraph}.E), hence the entropy with the island decreases with decreasing mass (Fig.\ref{threedimensionalgraph}.F). It can be seen that in other cases, by changing the values of the parameters, it is possible to achieve a negative derivative. The cyan surfaces are the surfaces of the critical mass $M>M_{cr}$, up to which the evaporation of a black hole can be investigated. The plots in the right column on Fig.\ref{threedimensionalgraph} show entropy with the island $S_{\cI}$ depending on the mass $M$ of the black hole at $\nu=0.5, 1, 1.5,4$ and various $b$. The plots Fig.\ref{threedimensionalgraph}.G, Fig.\ref{threedimensionalgraph}.H show that growth of entropy at $\nu=4$ (the special case of $\nu > \cV_2$) occurs in the region of masses less than the critical mass $M<M_{cr}$.
\newpage
\section{Time evolution under constraint equation}\label{timeevolution}
\subsection{Time evolution of mass and charge}

Until now, consideration of the evolution during the evaporation of a charged black hole has been carried out in terms of mass change, assuming that $M=M(t)$ is a monotonically decreasing function of time.
The explicit form of the function $M=M(t)$ was not fixed. In order to get  time dependence, it is sufficient to consider only one differential equation for $M$, since we assume that $Q$ evolves in time in accordance with the constraint equation \eqref{stateequation}.
\\

The evaporation of the Reissner-Nordström black holes is the subject of numerous consideration \cite{Gibbons-1875,Zaumen,Carter,Damour,
Page-1976,
Hiscock, Gabriel-2000,Sorkin:2001hf,
Ong:2019vnv}. In particular, the following  equation for the mass has been considered in \cite{Hiscock}
\bea\label{massdynamics}
\frac{dM}{dt}=  -\frac{\eta c \pi^2}{15} T^4  \sigma_0+\frac{Q}{r_+} \frac{dQ}{dt},
\eea
where first term is related with the emission of massless particles and corresponds to Stefan-Boltzmann law and the second term is related with mass loss rate due to the electromagnetic pair creation and corresponds to the first law of black hole thermodynamics. In \eqref{massdynamics} $\sigma_0$ is the geometrical optics cross section for the Reissner-Nordström black holes 
\bea
\label{sigma0}
\sigma_0 = \frac{\pi  \left(3 G M+\sqrt{9 G^2 M^2-8 G Q^2}\right)^4}{8 \left(3 G^2 M^2-2 G
	Q^2+G M \sqrt{9 G^2 M^2-8 G Q^2}\right)},
\eea
and $\eta \approx 2$ is an approximate constant depending on thermally averaged cross sections of the black hole for neutrinos, photons and gravitons.
In \cite{Hiscock} it is also assumed that  an extra equation takes place for $\frac{dQ}{dt}$. 
This is not the case in our consideration since we deal with the constraint equation \eqref{stateequation} and  the charge evolution  is given by
\bea
\frac{dQ}{dt} = \frac{\sqrt{G} \left(1-(1+\nu) \left(\frac{M}{\mu }\right)^{2 \nu
   }\right)}{\sqrt{1-\left(\frac{M}{\mu }\right)^{2 \nu }}}  \frac{dM}{dt}.
\eea
\\

\begin{figure}[h!]\centering
\includegraphics[width=65mm]{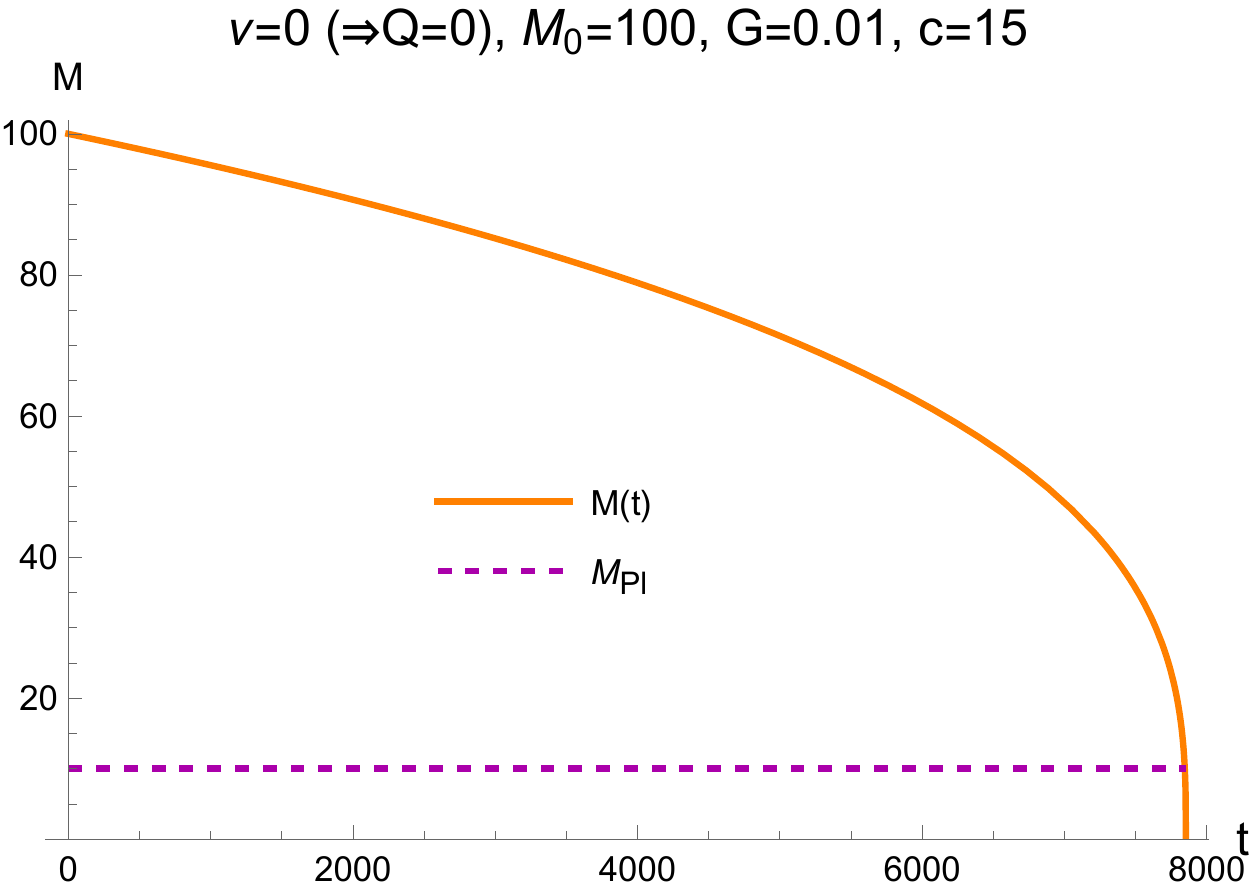}
	\qquad\includegraphics[width=65mm]{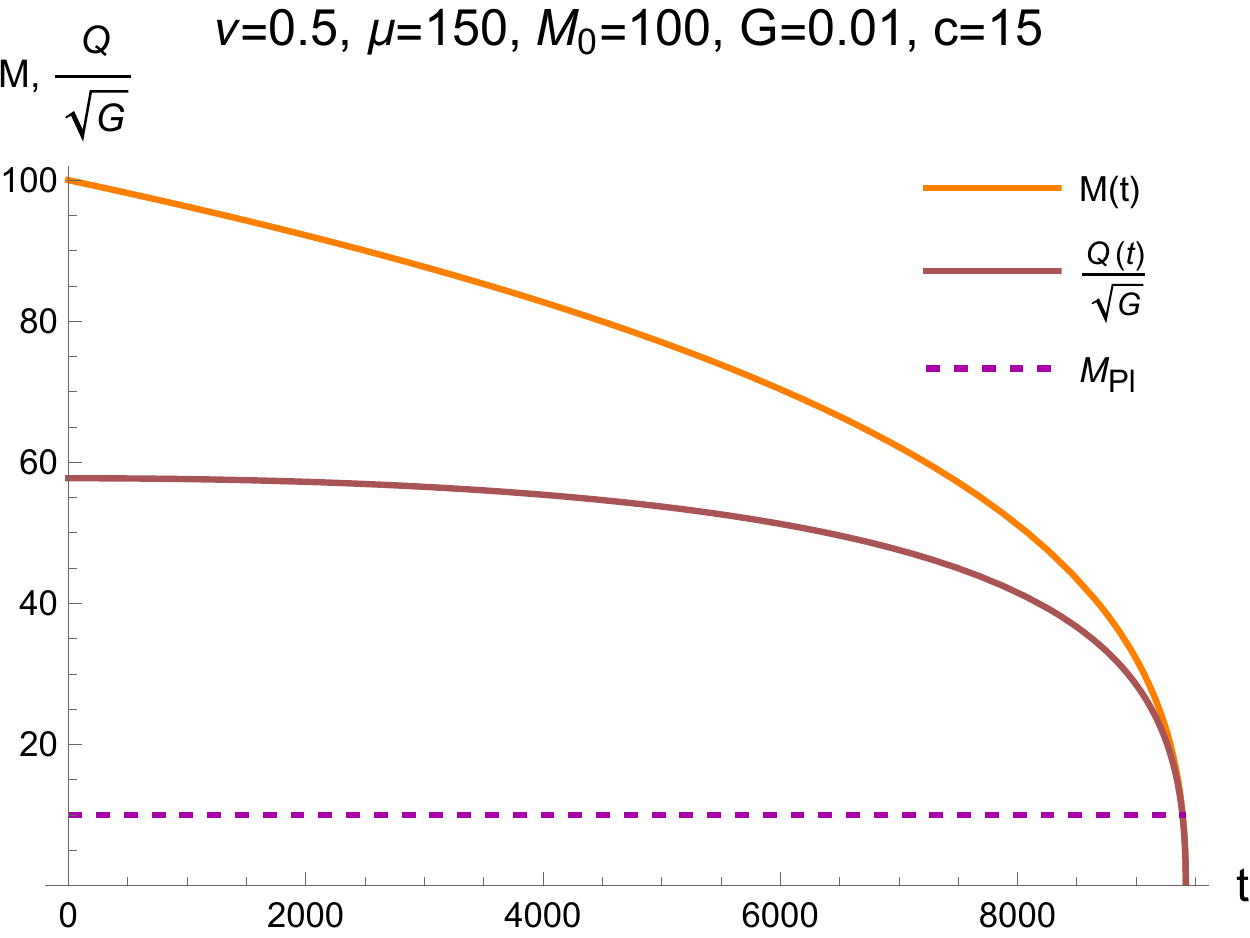}\\ \hspace{5mm} A) \hspace{70mm}B)\\
	\includegraphics[width=65mm]{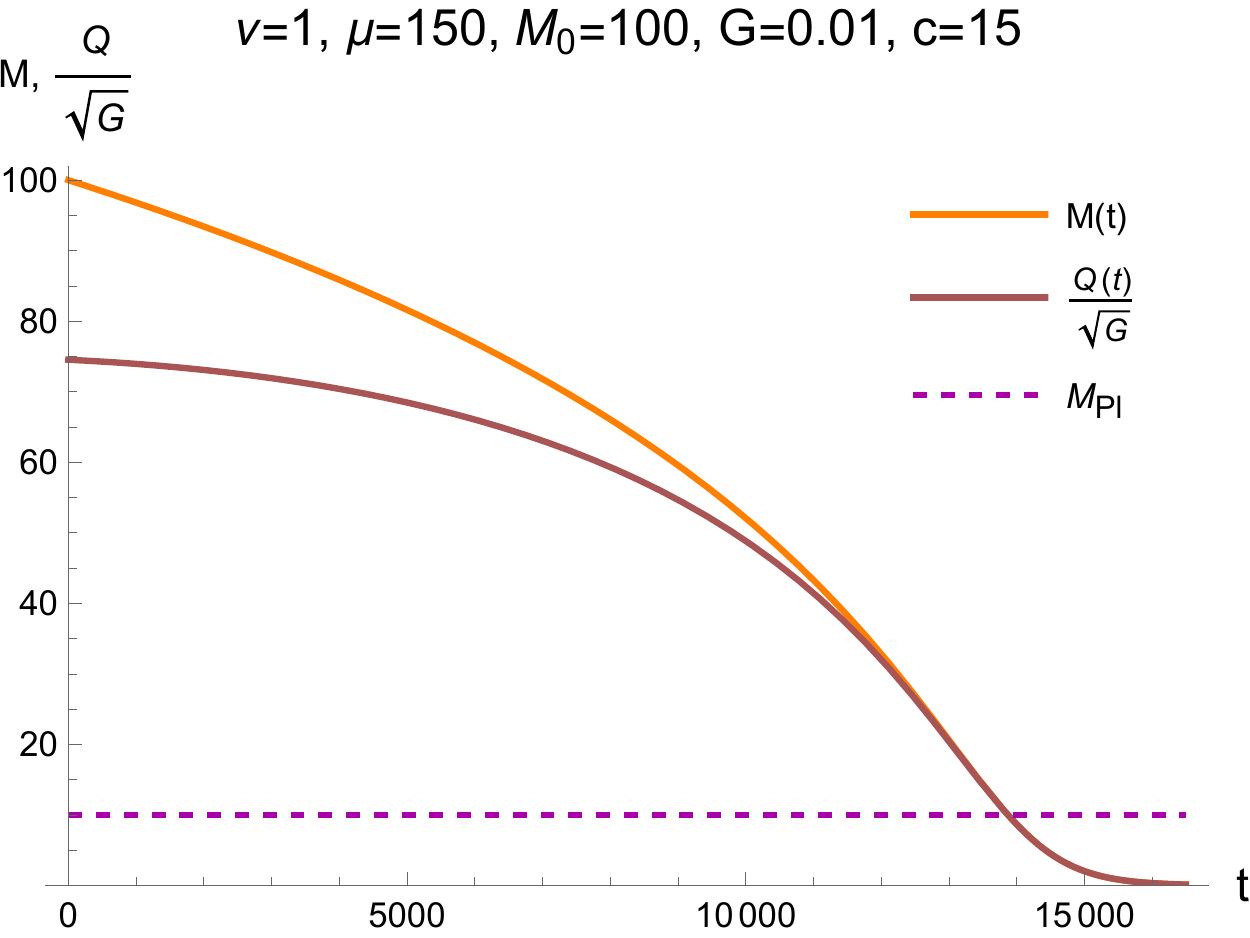}
	\qquad\includegraphics[width=65mm]{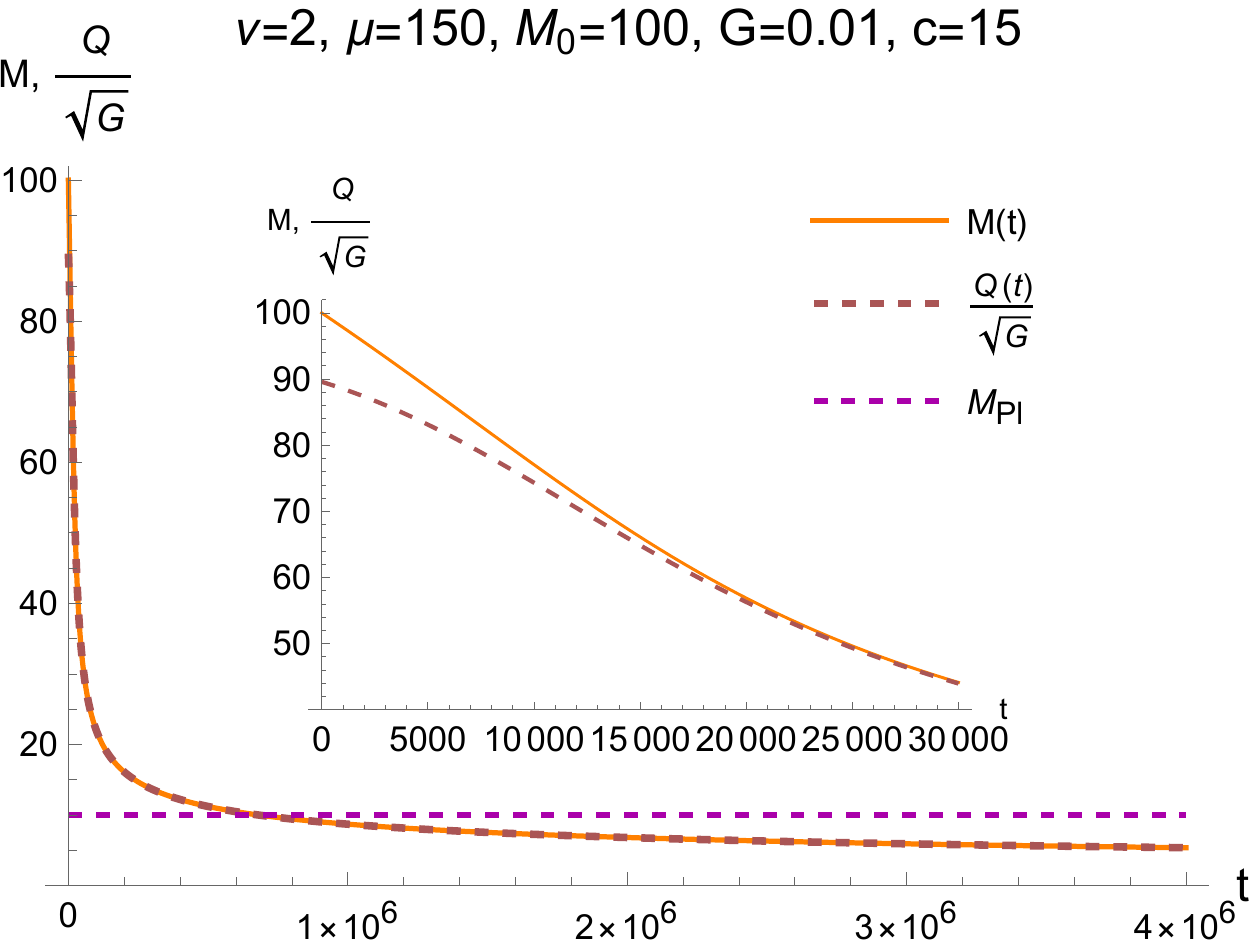} 	\begin{picture}(0,0)
\put(-409.9,16.5){\scriptsize{$M_{Pl}$}}
\put(-197,17.1){\scriptsize{$M_{Pl}$}}
\put(-408.5,177.0){\scriptsize{$M_{Pl}$}}
\put(-195.5,177.0){\scriptsize{$M_{Pl}$}}
\end{picture}\\ \hspace{5mm} C) \hspace{70mm}D)\\
	\caption{The dependence of mass $M(t)$ and charge $Q(t)$ (orange and brown lines, respectively) of a black hole on time from numerical research of the exact differential equation \eqref{massdynamics} at $\nu=0, 0.5, 1, 2$. At $\nu<1$ the evaporation time is finite, at $\nu \geq 1$ the evaporation time is infinitely large. The magenta dashed line corresponds to the Planck mass $M_{Pl} = 1/\sqrt{G}$.}% {\bf Math.file: Hiscock-and-our-EOS.nb}}
	\label{massandchargetime}
\end{figure}
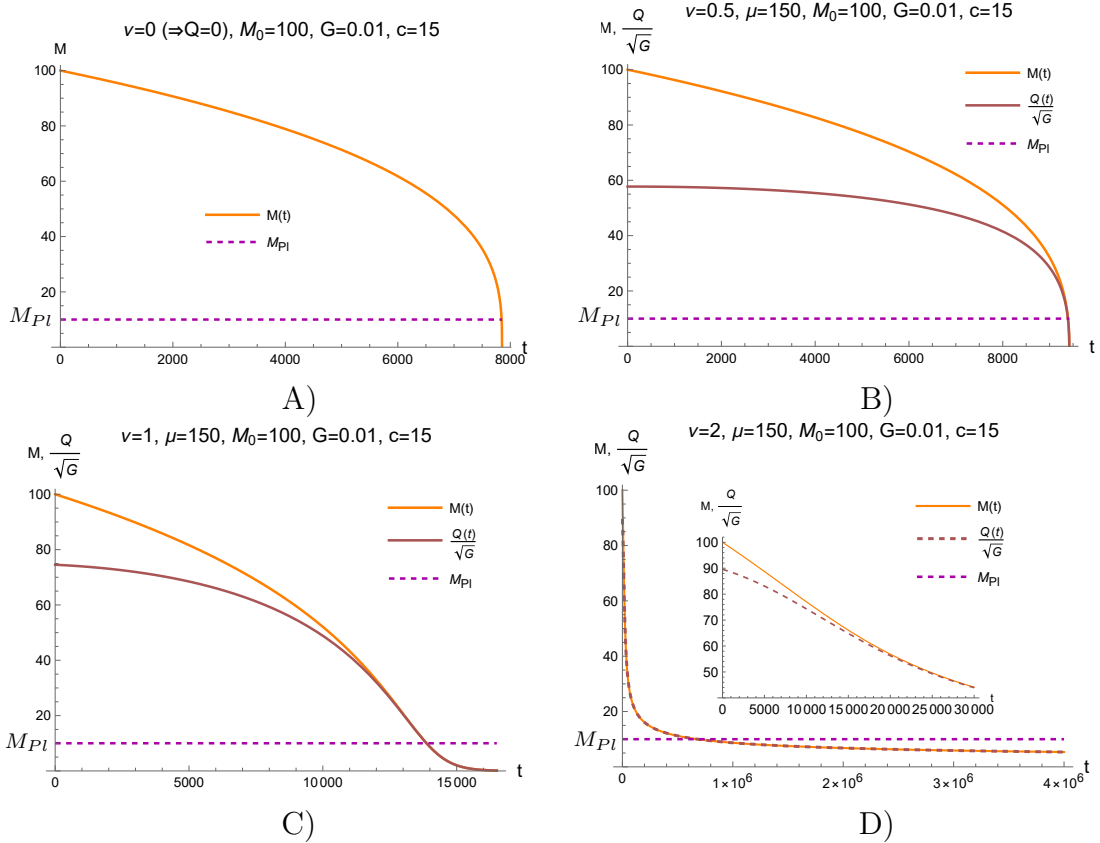

Therefore, equation \eqref{massdynamics} under constraint equation \eqref{stateequation} becomes
\bea\label{diffur1}
\frac{dM}{dt}=-\frac{\eta c \left(3+\sqrt{1+8 \left(\frac{M}{\mu }\right)^{2 \nu }}\right)^4
  M^{3 \nu -2}}{1920 \pi  G^2 \mu^{3 \nu } \left(1+\left(\frac{M}{\mu
   }\right)^{\nu }\right)^7 \left(1+2 \left(\frac{M}{\mu }\right)^{2 \nu }+\sqrt{1+8
   \left(\frac{M}{\mu }\right)^{2 \nu }}\right) \left(1+(\nu +1)
   \left(\frac{M}{\mu }\right)^{\nu }\right)}. \nn \\
\eea
Let us expand the right side of (\ref{diffur1}) in powers of $\left(\frac{M}{\mu }\right)^{ \nu } \ll 1$.
Then (\ref{diffur1}) in the leading order is 
\bea\label{approxequation}
\frac{dM}{dt}=-A \,M^{3 \nu -2}, \quad \mbox{where} \quad A = \frac{\eta c}{15 \pi  G^2 \mu^{3 \nu } }.
\eea
The solution of \eqref{approxequation} with the initial condition  $M(0)=M_0$ is
\be\label{solutionofequation}
M(t) =\left\{\begin{array}{cc}
    M_0 \,\left( 1-\cfrac{3(1-\nu) A t}{M^{3(1-\nu)}_0}\right)^{\frac{1}{3(1-\nu)}}, & \nu \neq 1,\\\,&\\
 M_0 \,e^{-A t}, &  \nu=1.
\end{array}\right.
\ee
Fig.\ref{massandchargetime} shows the time dependence of the mass $M$ and charge $Q$ of a black hole for various $\nu$. From \eqref{solutionofequation} at $\nu<1$ the evaporation time $M(t_{evap})=0$ is
\be\label{evaporationtime}
t_{evap} = \frac{M^{3(1-\nu)}_0}{3(1-\nu) A}.
\ee
However, for $\nu \geq 1$, the total evaporation time is infinitely long.

\subsection{Time evolution of entanglement entropy}

Up to this point, only the entanglement entropy of configuration with  an island (\ref{entropwithisland}) has been investigated. With the time dependence of the mass \eqref{solutionofequation}, it is now possible to study entropy without an island \eqref{withoutislandconstraint}. Expanding \eqref{withoutislandconstraint} over small $\alpha \ll 1$ and $\frac{GM}{b} \ll 1$, we get
at $\frac{t  M^{\nu -1}}{G \mu ^{\nu }} \gg 1$
\be\label{withoutisland2}
S_{n\cI} \simeq \frac{c}{3} \frac{t  M^{\nu -1}}{G \mu ^{\nu }}+\frac{c}{3} \log \Big[2 \sqrt{G}  M^{1-\nu }  \mu ^{\nu } \Big].
\ee
From \eqref{solutionofequation} it can be seen that the entropy without an island \eqref{withoutisland2} increases monotonically with time for any $\nu \geq 0$.
\\

One can compare the entropy with and without an island and study which configuration dominates. At early times the growing over time entropy without an island \eqref{withoutislandconstraint} dominates, but at the moment of its intersection  with the entropy with the island (\ref{entropwithisland}), $S_{n\cI} (t) = S_{\cI} (t)$, the latter begins to dominate. 
Thus, it is the behavior of the entropy with the island that is responsible for the decrease or increase in the entanglement entropy at the end of the black hole evaporation.
\\

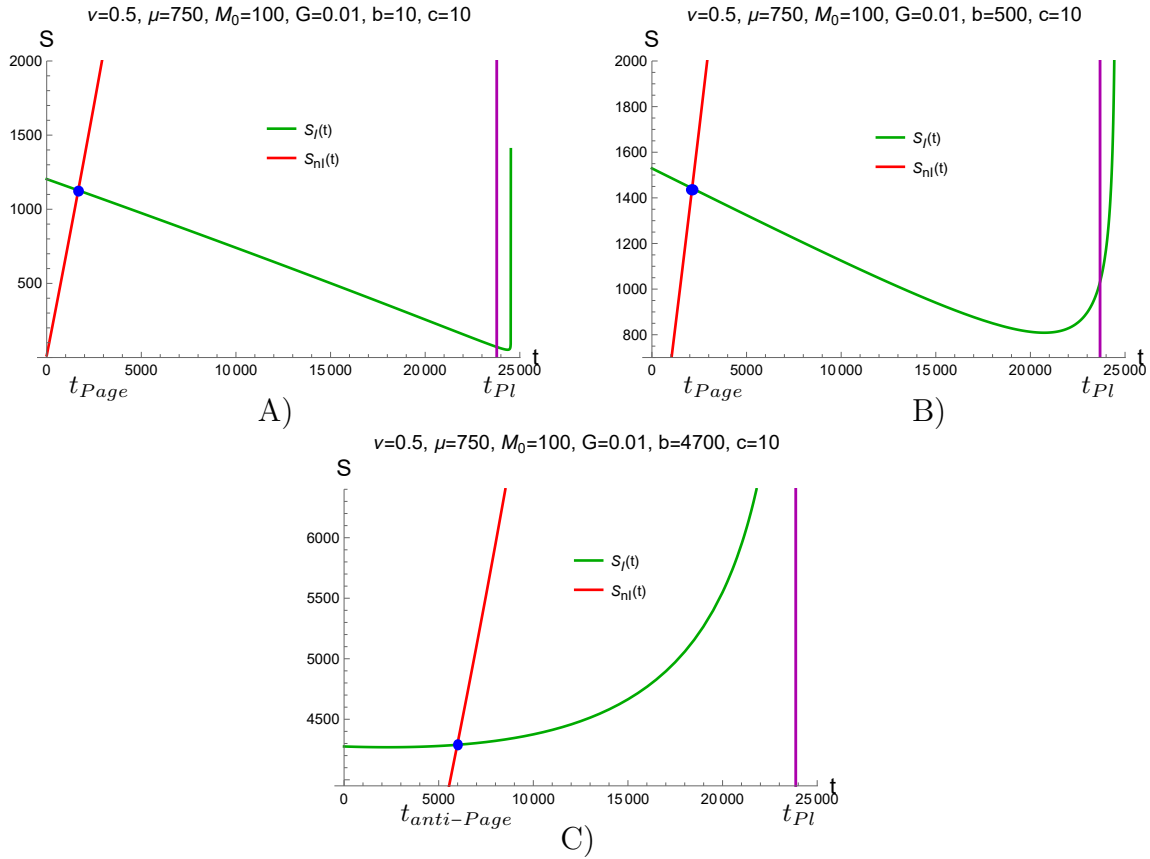
\begin{figure}[h!]\centering
\includegraphics[width=70mm]{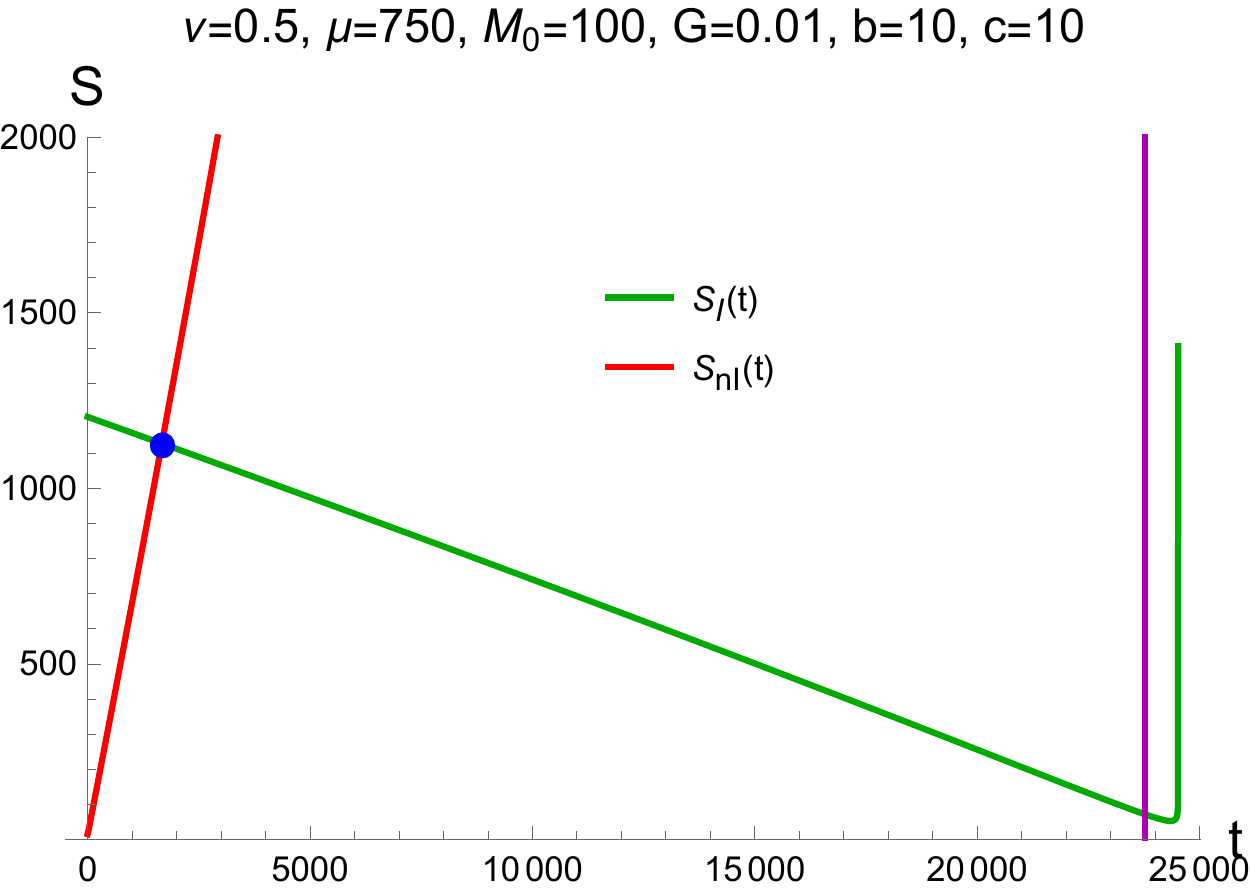}
	\qquad\includegraphics[width=70mm]{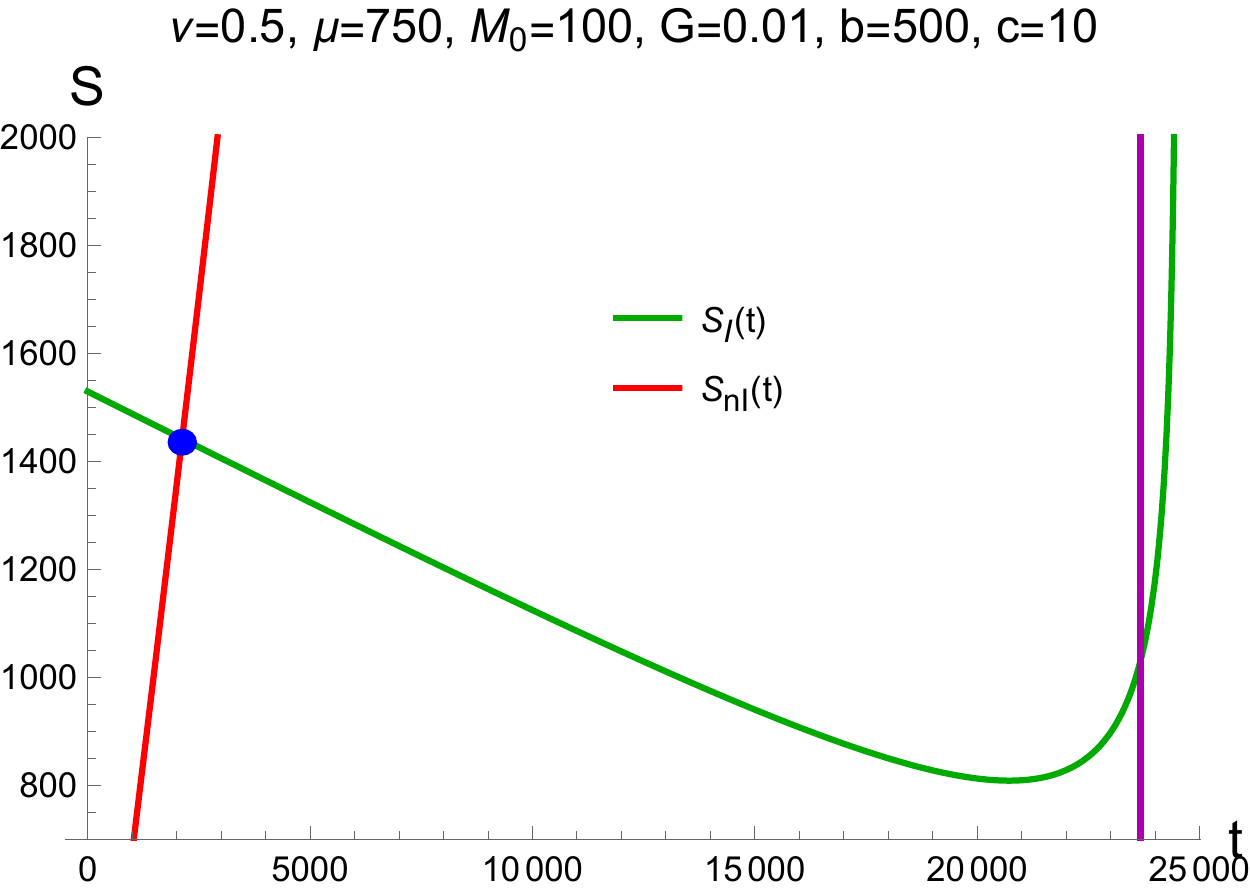} \\ \hspace{5mm} A) \hspace{80mm}B) \\
\includegraphics[width=70mm]{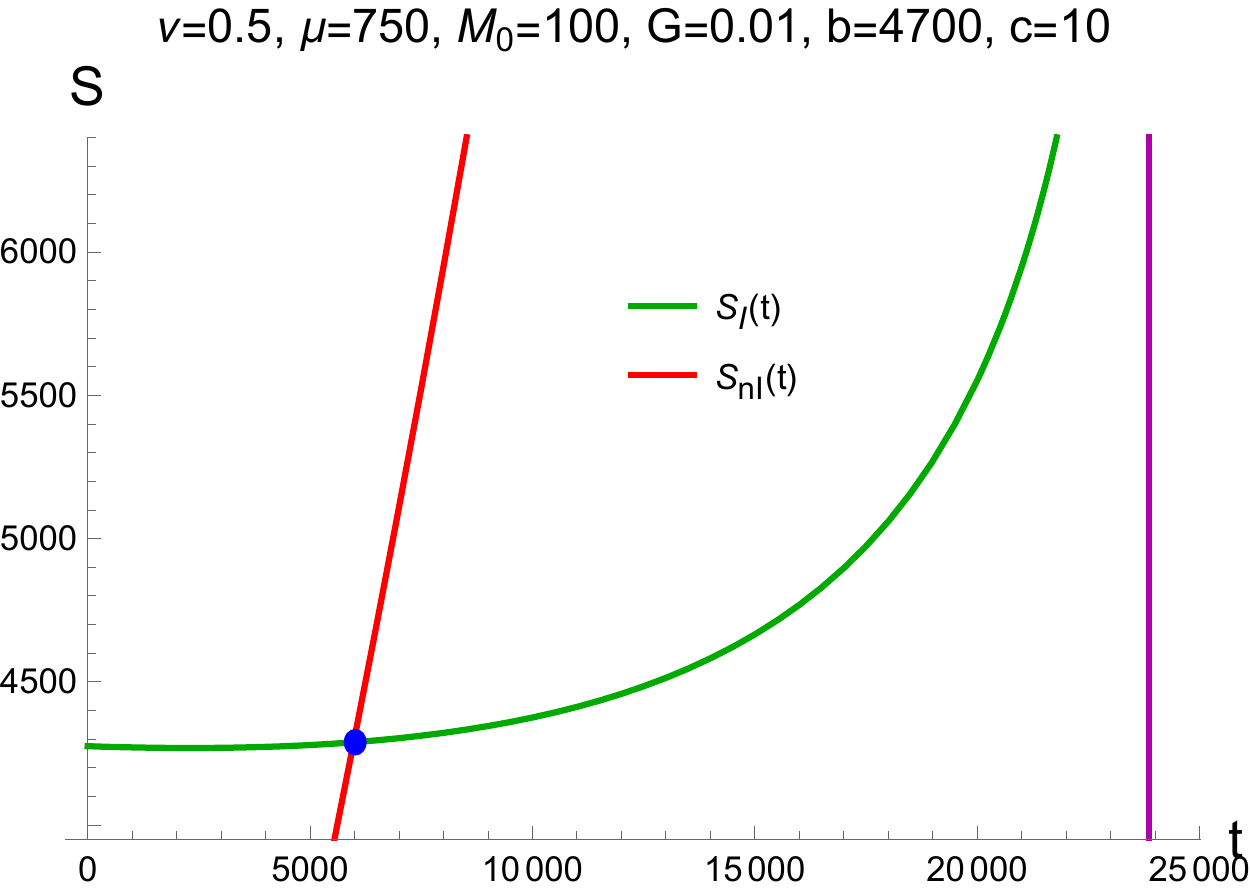}	
	\begin{picture}(0,0)
\put(-168,-5.5){\footnotesize{$t_{anti-Page}$}}
\put(-25,-5.5){\footnotesize{$t_{Pl}$}}
\put(-292,156){\footnotesize{$t_{Page}$}}
\put(-137,156){\footnotesize{$t_{Pl}$}}
\put(-62,156){\footnotesize{$t_{Page}$}}
\put(87,156){\footnotesize{$t_{Pl}$}}
\end{picture}\\
C)
	\caption{Dependence of entropy with $S_{\cI}(t)$ and without island $S_{n\cI}(t)$ (green and red lines, respectively) on time at $\nu=0.5$ for various $b$. The blue circles correspond to the time when the entropies intersect $S_{\cI} (t) = S_{n\cI} (t)$. The plots A), B) show that at this moment the entropy with an island decreases (Page time) and the plot C) shows that the entropy with island increases (anti-Page time), which is achieved by increasing $b$. The magenta line corresponds to $t_{Pl}$, $M(t_{Pl}) = M_{Pl}$, which for a given ratio of parameters is approximately equal to the critical time $t_{cr}$. The plot A) corresponds to the absence of growth of entropy with an island at $t<t_{cr}$. The plots B), C) correspond to the appearance of growth of entropy with an island at $t<t_{cr}$ with increasing $b$.}% {\bf Math.file: Hiscock-and-our-EOS.nb}}
	\label{timedep1}
\end{figure}

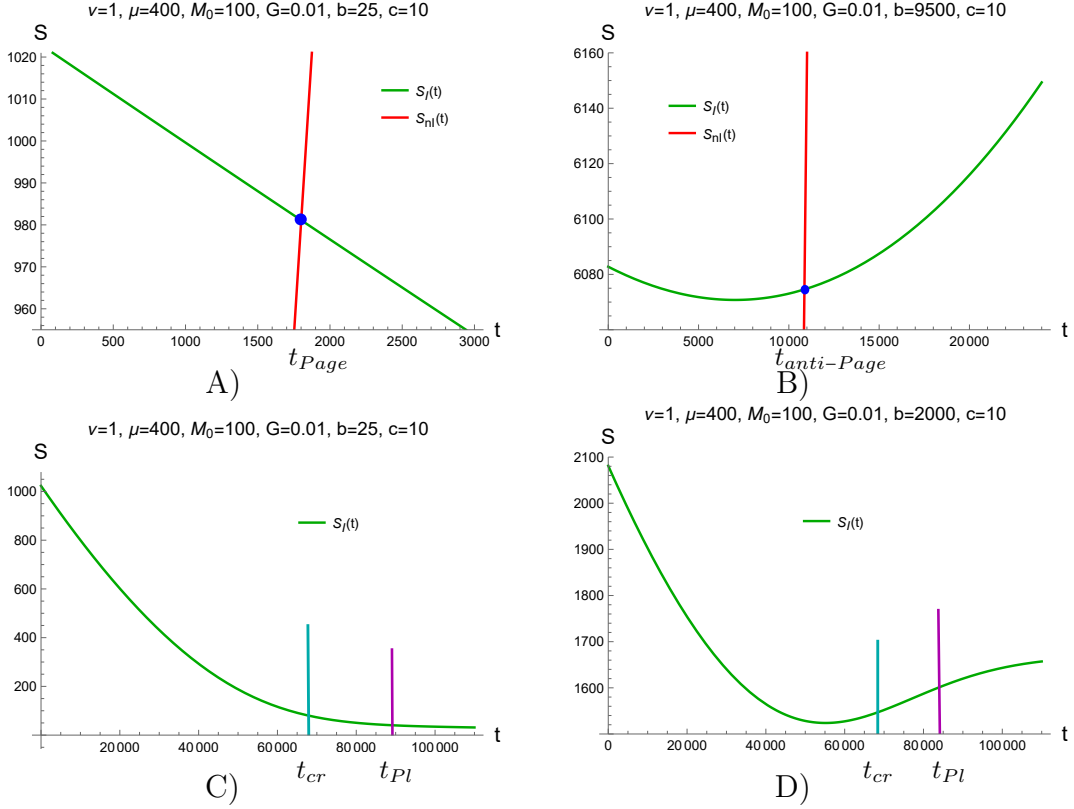
\begin{figure}[h!]\centering
\includegraphics[width=65mm]{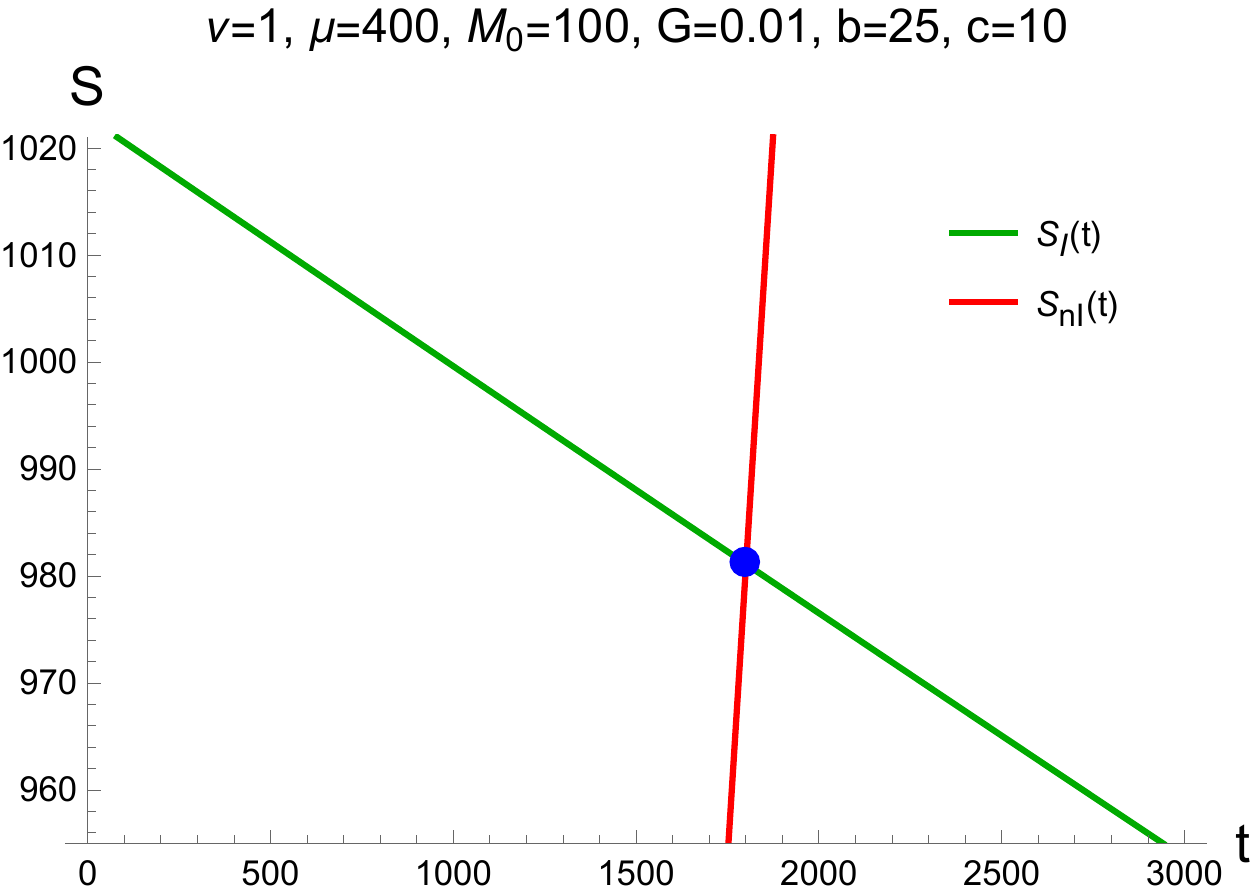}
	\qquad\includegraphics[width=65mm]{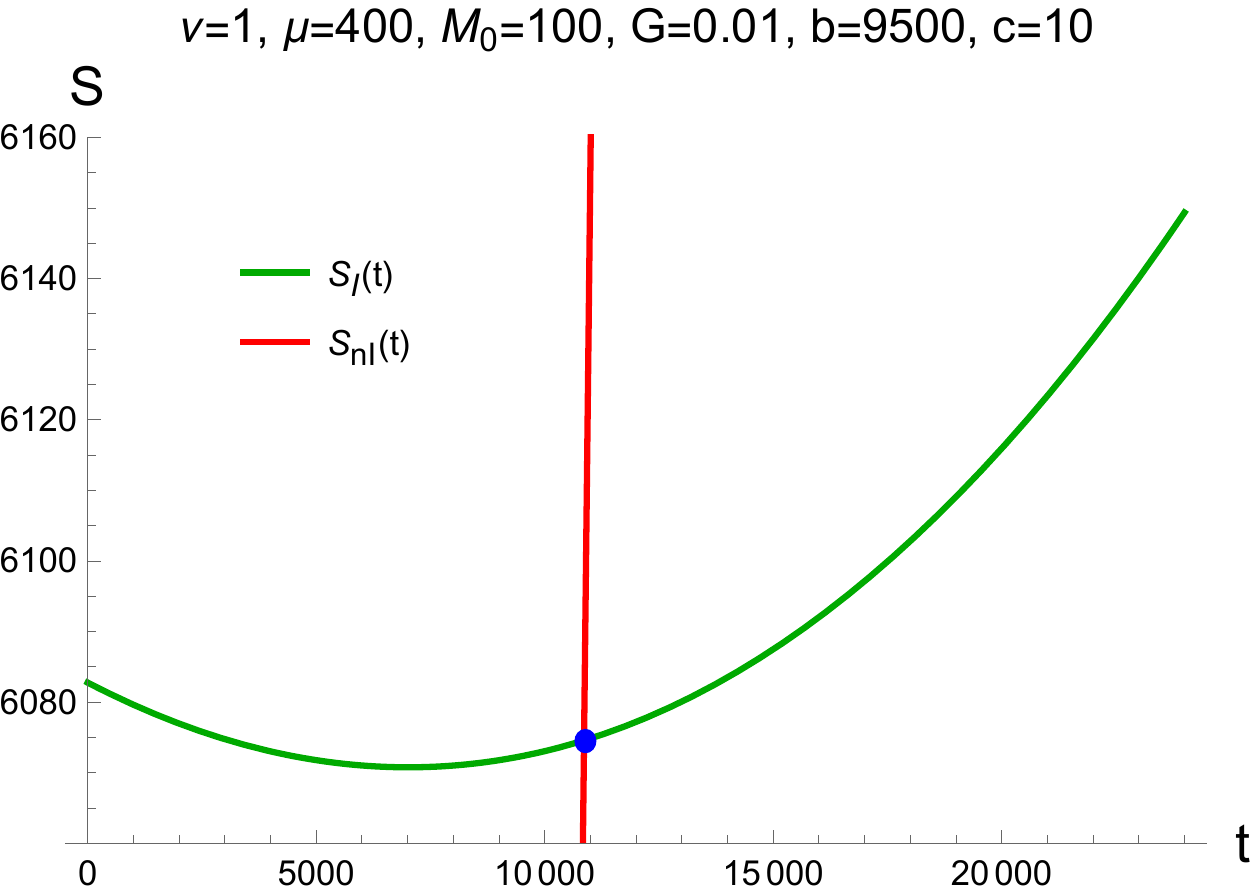}\\ \hspace{-9mm}  A) \hspace{69mm}B)\\
	\includegraphics[width=65mm]{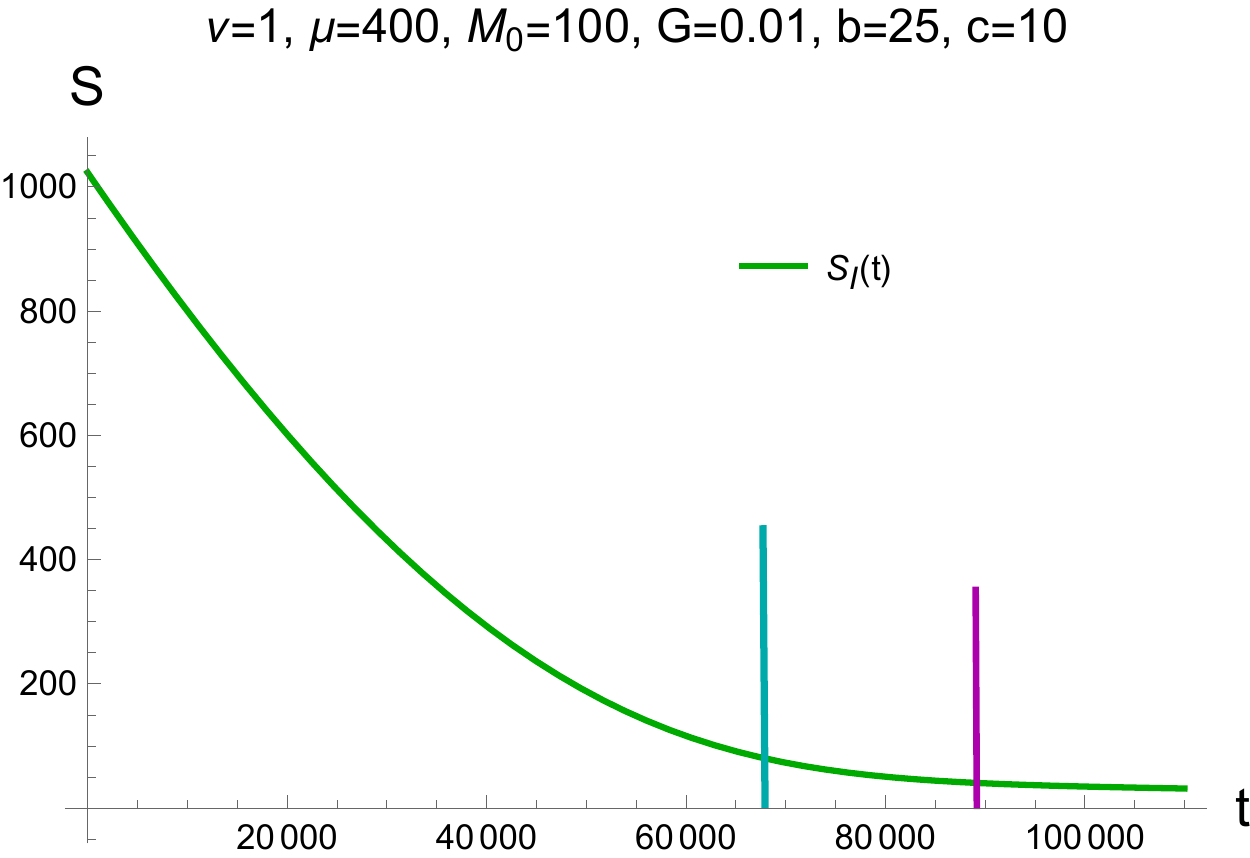}
	\qquad\includegraphics[width=65mm]{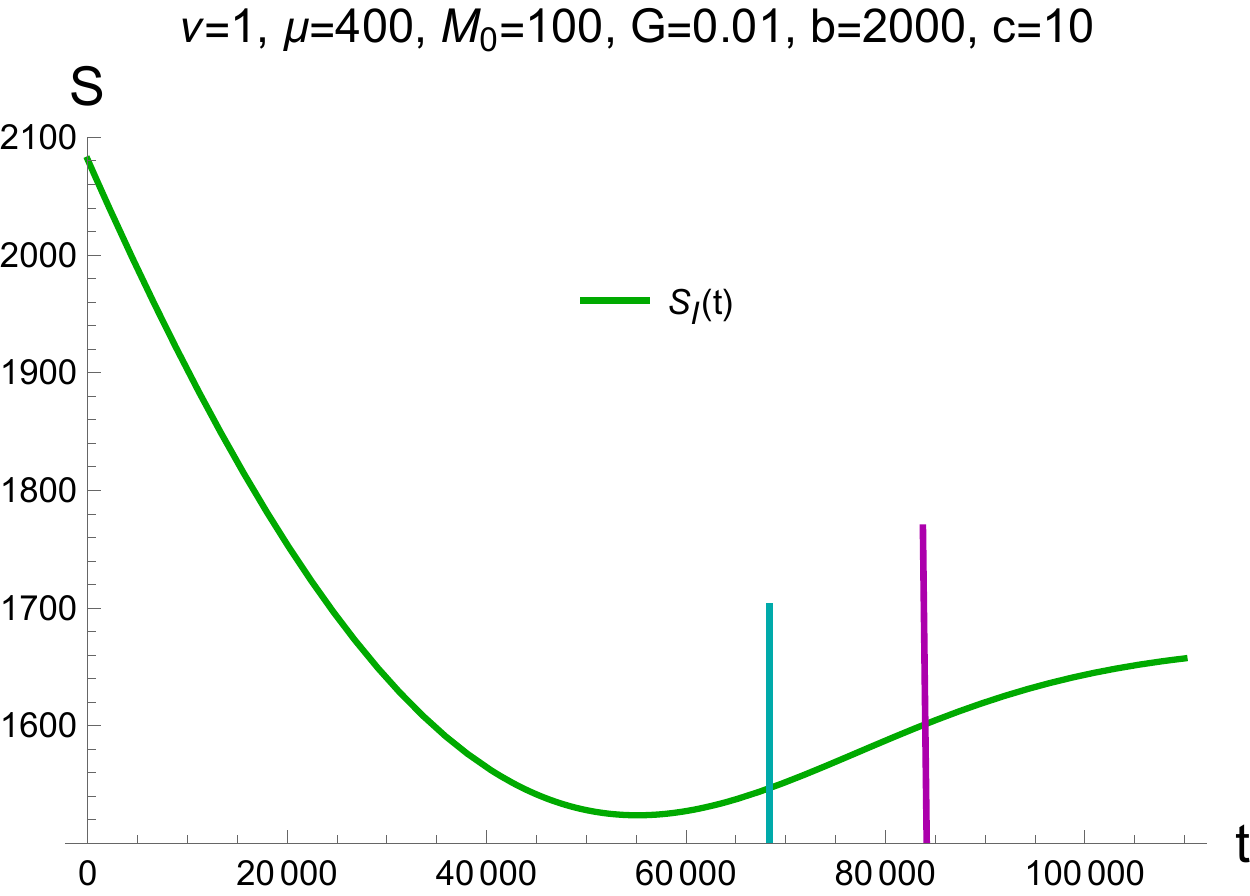} \\ \hspace{-9mm}  C) \hspace{69mm}D)\\
	\begin{picture}(0,0)
\put(89,179){\footnotesize{$t_{anti-Page}$}}
\put(-93,179){\footnotesize{$t_{Page}$}}
\put(122,23){\footnotesize{$t_{cr}$}}
\put(148,23){\footnotesize{$t_{Pl}$}}
\put(-90,23){\footnotesize{$t_{cr}$}}
\put(-59,23){\footnotesize{$t_{Pl}$}}
\end{picture}
	\caption{Dependence of entropy with $S_{\cI}(t)$ and without island $S_{n\cI}(t)$ (green and red lines, respectively) on time at $\nu=1$ for various $b$. The blue circles on A) and B) correspond to the time when the entropies intersect $S_{\cI} (t) = S_{n\cI} (t)$. The plot A) shows that at this moment the entropy with island decreases (Page time) and the plot B) shows that the entropy with island increases (anti-Page time), which is achieved by increasing $b$. On plots C) and D) the cyan and magenta lines correspond to $t_{cr}$, $M(t_{cr}) = M_{cr}$ and $t_{Pl}$, $M(t_{Pl}) = M_{Pl}$, respectively.   The plot C) shows the decrease in entropy with island at $t<t_{cr}$. The plot D) shows that as $b$ increases, the entropy with island starts to increase at $t<t_{cr}$.}% {\bf Math.file: Hiscock-and-our-EOS.nb}}
	\label{timedep2}
\end{figure}

\begin{figure}[h!]\centering
\includegraphics[width=70mm]{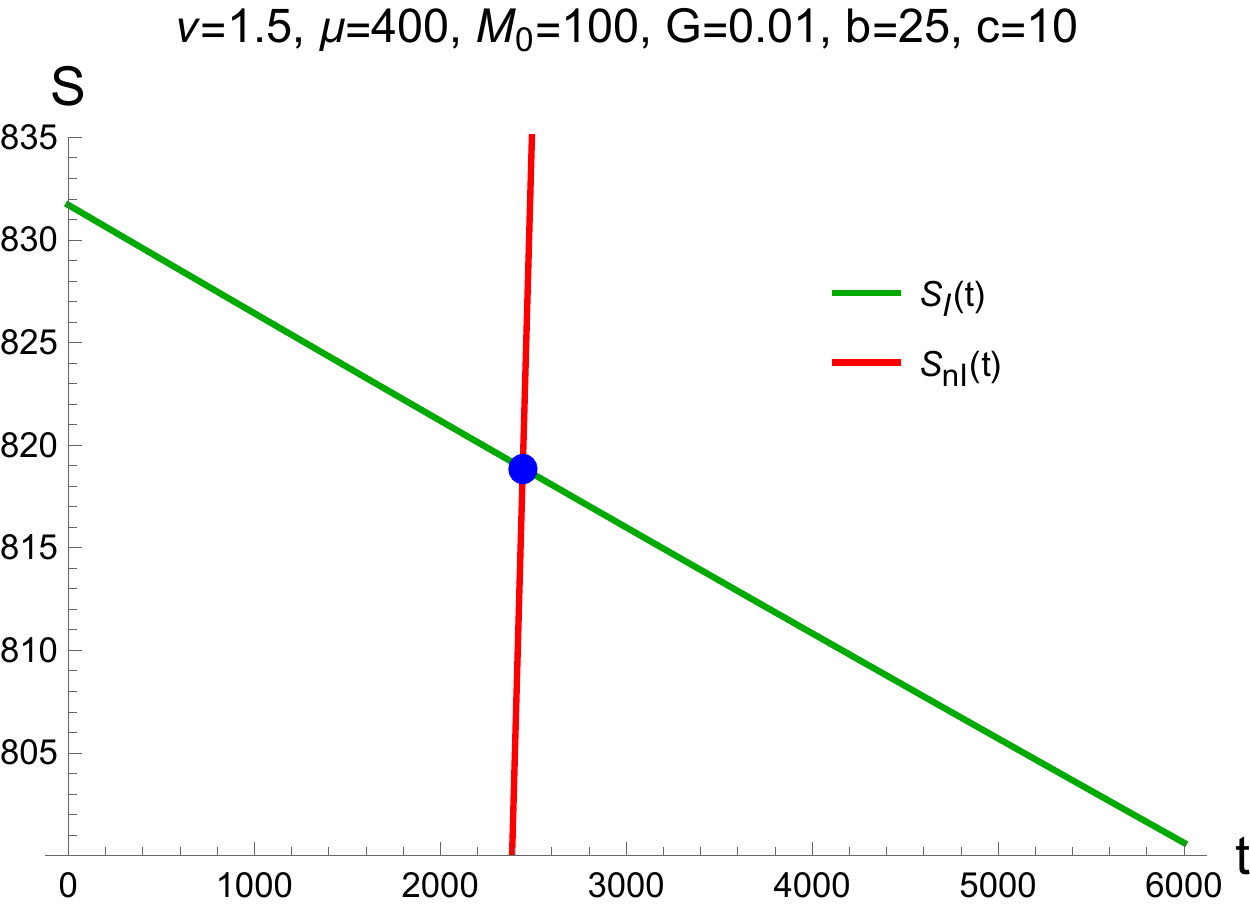}
	\qquad\includegraphics[width=70mm]{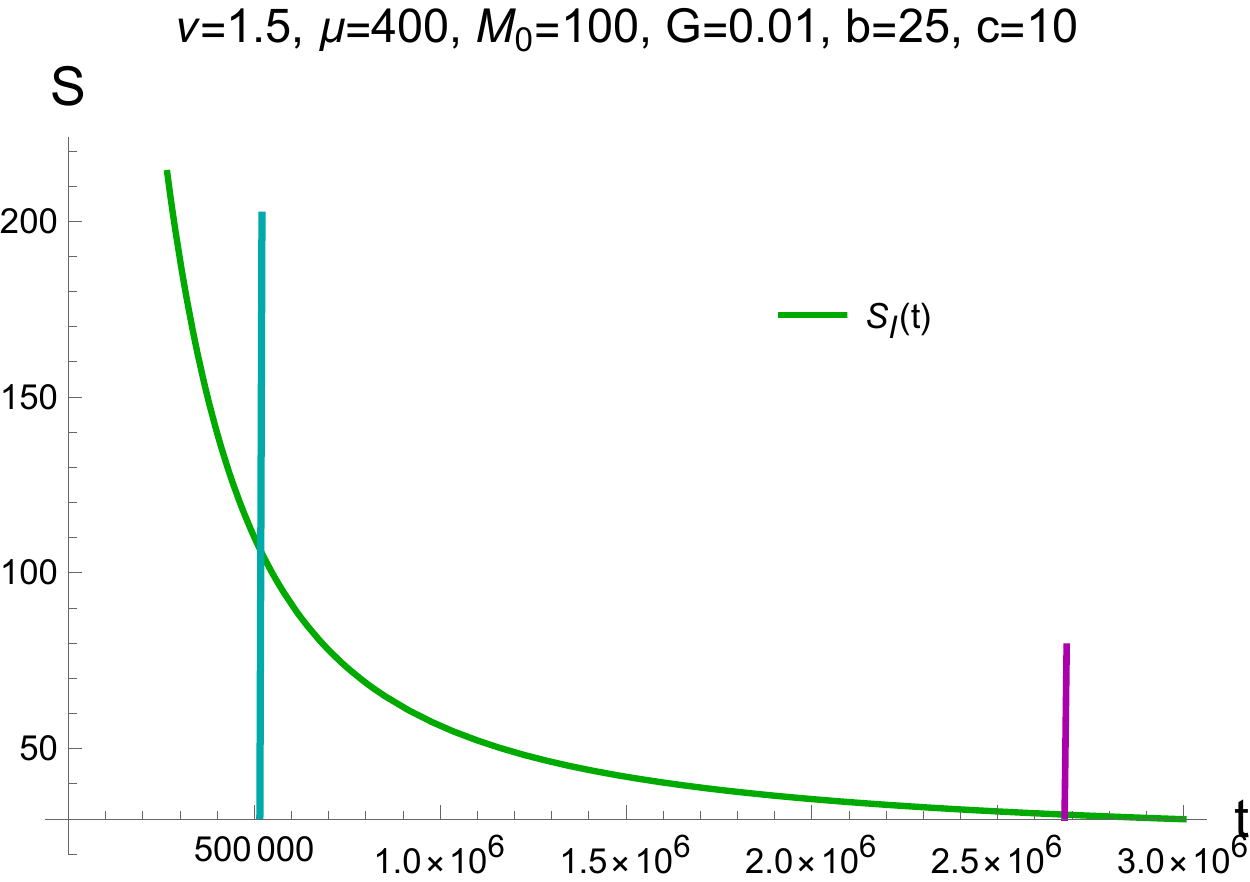}
		\begin{picture}(0,0)
\put(-166,-8.5){\footnotesize{$t_{cr}$}}
\put(-38,-8.5){\footnotesize{$t_{Pl}$}}
\put(-349,-8.5){\footnotesize{$t_{Page}$}}
\end{picture}\\
\hspace{-15mm} A) \hspace{85mm}B)
	\caption{Dependence of entropy with $S_{\cI}(t)$ and without island $S_{n\cI}(t)$ (green and red lines, respectively) on time at $\nu=1.5$. The blue circle on A) corresponds to the time when the entropies intersect $S_{\cI} (t) = S_{n\cI} (t)$. The plot A) shows that at this moment the entropy with the island decreases (Page time). On plot B) the cyan and magenta lines correspond to $t_{cr}$, $M(t_{cr}) = M_{cr}$ and $t_{Pl}$, $M(t_{Pl}) = M_{Pl}$, respectively. The plot B) shows the decrease in entropy with island at $t<t_{cr}$.}% {\bf Math.file: Hiscock-and-our-EOS.nb}}
	\label{timedep3}
\end{figure}
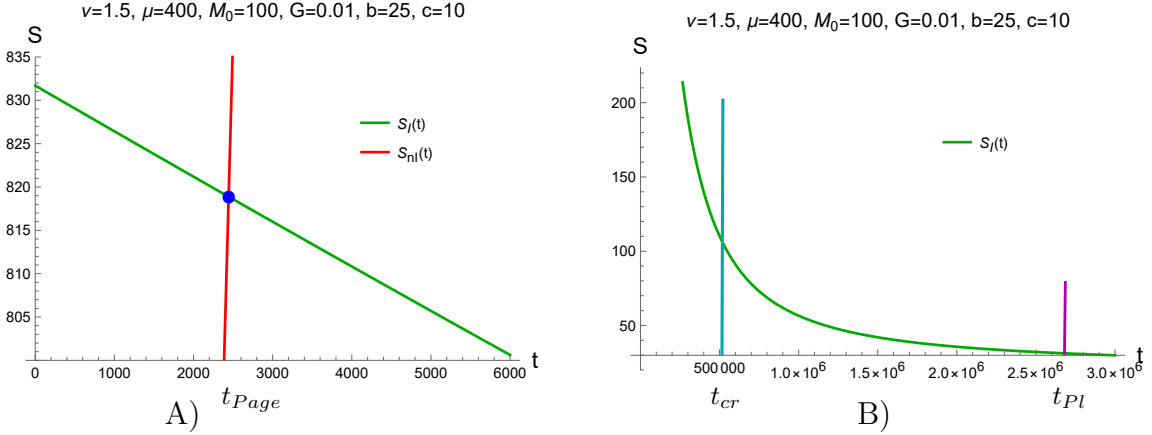

In Appendix  \ref{appendixapproximation} inequalities that give limitation on the used approximations to  the entropy with island are considered. There we show that there is the
critical mass $M_{cr}$, above which we can use the approximation for the entanglement entropy with an island, given by the formula \eqref{islandch}.
This mass corresponds to  $t_{cr}$,
$M(t_{cr}) = M_{cr}$, up to which  the evaporation process can be considered in given approximation.  In Fig.\ref{timedep2}, Fig.\ref{timedep3} the critical time $t_{cr}$ corresponds to the vertical cyan lines. Also in Fig.\ref{timedep1}, Fig.\ref{timedep2}, Fig.\ref{timedep3}, the magenta lines indicate the time $t_{Pl}$, $M(t_{Pl}) =  M_{Pl}$. Depending on the parameters, $M_{cr}$ can be either more or less than $M_{Pl}$.
\\

At $\nu<1$ at the end of the evaporation ($M \to 0$) the entropy with island explodes \eqref{entropyasymp} and the black hole evaporates in a finite time \eqref{evaporationtime}. In Fig.\ref{timedep1} dependencies of entropy with and without island on time at $\nu=0.5$ for various $b$ are shown. Depending on the parameter $b$ at the moment of intersection of the entropy without and with island, the latter can either decrease (Fig.\ref{timedep1}.A) or increase (Fig.\ref{timedep1}.B), which corresponds to the so-called Page and anti-Page times, respectively. Thus, the time evolution of the entanglement entropy at $\nu<1$ resembles the Schwarzschild case \cite{Arefeva:2021kfx}.

$$\,$$
\\

Let us consider the domain $\nu \in [1,\cV_1)$, in which, for  a certain choice of model parameters
an increase of entropy of configurations  with an island over time can occur. 
In Fig.\ref{timedep2} dependencies of entropy with and without island on time at $\nu=1$ for various $b$ are shown. Depending on the parameter $b$ Page (Fig.\ref{timedep2}.A) or anti-Page (Fig.\ref{timedep2}.B) times can be realized. The difference from the Schwarzschild case \cite{Arefeva:2021kfx} is that the total evaporation time is infinitely long.
\\

Let us consider the domain $\nu \in [\cV_1, \cV_2]$. In Fig.\ref{timedep3} dependencies of entropy with and without island on time at $\nu=1.5$ which almost certainly belongs to $[\cV_1, \cV_2]$ are shown. It was demonstrated in Section~\ref{analysis} that in this case the entropy with island does not increase for the entire mass range, i.e. at all times. Thus, the Page time (Fig.\ref{timedep3}.A) and the infinitely long decrease in entropy with island (Fig.\ref{timedep3}.B) are realized.
\\

We noted in Section \ref{analysis} that it seems for $\nu > \cV_2$ the entropy with an island increases only in the mass region less than the critical mass $M<M_{cr}$. Thus, the Page time and the further decrease in entropy at $t<t_{cr}$ are realized.

\section{Conclusion and discussion}
We have considered the black hole mass dependence  of the entanglement entropy of the Hawking radiation obtained by the island recipe for the asymptotically flat non-extremal Reissner-Nordström black hole  
under the certain class of constraints \eqref{stateequation}.
The entanglement entropy of radiation of the eternal Reissner-Nordström
 black hole was considered previously in \cite{Kim:2021gzd, Wang:2021woy}. The specific of our consideration is in the assumption of the special constraint \eqref{stateequation}. This constraint
 permits to consider a near-extremal case and avoid the explosion of the temperature at the end of evaporation of the Reissner-Nordström black hole.
\\

We have investigated how close to the extremal case it is possible to approach so that approximations under which the entropy with an island is derived remain satisfied.
We have found the critical mass $M_{cr}$ so that for $M>M_{cr}$ assumptions necessary to derive the approximated value of the entanglement entropy with an island are satisfied.
\\

The behavior of the entanglement entropy essentially depends on the specific relationship between the charge and mass of the Reissner-Nordström black hole. It is shown that if parameter $\nu$ specifying  the constraint equation \eqref{stateequation}, belongs the interval  $\cV_1 < \nu <\cV_2$, where  $1<\cV_1$ and $\cV_2<2$, the entanglement entropy with the island configuration decreases with decreasing mass for any physically meaningful  values of the parameters. Apparently, at $\nu>\cV_2$ the growth of entropy with the island occurs at masses  less than the critical mass, i.e. $M < M_{cr}$. Thus, the introduction of a charge makes it possible to avoid an increase in the entanglement entropy with an island configuration at small masses, which occurs for the Schwarzschild black holes \cite{Arefeva:2021kfx}.
\\

Under the constraint equation we have considered the time differential equation  for the mass and found the time evolution of the mass $M=M(t)$ and charge $Q=Q(t)$ of the black hole. We assume that  under the assumption that the mass and the charge of the black hole vary slowly with time we can use the same formula for entanglement entropy for evaporating black hole as for the eternal Reissner-Nordström black hole.   Knowledge of the explicit time dependence makes it possible to investigate whether entropy with or without an island dominates at a given time. At early times, the growing over time entropy without an island dominates, but at the moment of its intersection with the entropy with an island the latter begins to dominate. For special values of parameter $\nu$, $\nu<\cV_1$, and depending on others  model parameters the entropies with and without an island can intersect at the moment when the entropy with the island decreases (Page time) or increases (anti-Page time), which is similar to the situation for Schwarzschild black holes \cite{Arefeva:2021kfx}.
\\

\section*{Acknowledgements}
The work of I.A. and T.R. is supported by the Russian Science Foundation (project 20-12-00200, Steklov Mathematical Institute).

%%%%%%%%%%%%%%%%%%%%%%%%%%%%%%%%%%%%%%%%%%%%%%%%%%%
\appendix

%%%%%%%%%%%%%%%%%%%%%%%%%%%%%%%%%%%
\section{Approximations for entanglement entropy near extremal regime}\label{appendixapproximation}

 In this appendix we study how close to the extremal case   one can get so that approximations  under  which  the entropy  with an  island \eqref{islandch}  are  derived remain valid. 
Our consideration will show that for given parameters $b,c,G,\mu,\nu$
there exists $M_{cr}(b,c,G,\mu,\nu)$ such that for $M>M_{cr}$ all approximations are valid.
\\

For this purpose let us sketch the main steps of derivation of the entropy  with the  island \eqref{islandch}. From the general formula \eqref{GEformula3} one gets the entropy with the island for the configuration presented in Fig.\ref{fig:island}
\bea\label{EEI}
		S_{\cI} (a, t_a) &= &\frac{2\pi a^2}{G} + \frac{c}{6}\log \left(\frac{16f(a)f(b)}{\kappa^4}\cosh^2\kappa t_a \cosh^2\kappa t_b \right)\\
		&+& \frac{c}{3}\log\left| \frac{\cosh\kappa(r_\ast(a)-r_\ast(b))-\cosh\kappa(t_a-t_b)}{\cosh\kappa(r_\ast(a)-r_\ast(b))+\cosh\kappa(t_a+t_b)} \right|. \nn
\eea
The following notations in \eqref{EEI} are used
\bea
r_* (r) = r+ \frac{r_+^2}{r_+-r_-}\log \Big|\frac{r-r_+}{r_+}\Big|-\frac{r_-^2}{r_+-r_-}\log\Big|\frac{r-r_-}{r_-}\Big|, \quad \kappa = \frac{r_+-r_-}{2 r_+}.
\eea
It is assumed that the entropy \eqref{EEI} is extremized with respect to the coordinates $(a, t_a)$.
\\

Assuming that $a=r_+ + X$, $X \ll r_+$, the following approximation is used
\be\label{approxcosh}
\cosh\kappa(r_\ast(a)-r_\ast(b)) \simeq \frac{1}{2} e^{\kappa(r_\ast(b)-r_\ast(a))} = \frac{1}{2} e^{\kappa (b-a)} \left( \frac{b-r_+}{a-r_+} \right)^{\frac{1}{2}} \left( \frac{a-r_-}{b-r_-} \right)^{\frac{r^2_-}{2 r^2_+}},
\ee
that is is satisfied if
\be\label{coshinus}
e^{2\kappa(r_\ast(b)-r_\ast(a))} \gg 1.
\ee
The late time approximation is also employed
\be\label{lta}
\cosh \kappa (t_a+t_b) \gg \frac{1}{2} e^{\kappa(r_\ast(b)-r_\ast(a))}, \quad \kappa\, t_{a,b} \gg 1.
\ee
Then it can be seen that extremization over $t_a$ gives $t_a = t_b$.
\\

The following expansion is carried out
\be\label{firstexpan}
\log \left[1-2e^{\kappa(r_\ast(a)-r_\ast(b))} \right] \simeq -2 e^{\kappa (r_*(a)-r_*(b))},
\ee
that is true if the following inequality holds
\bea\label{approx1}
{\mathbf Y}_1: \, \, 2 e^{\kappa (r_*(a)-r_*(b))}  \ll 1.
\eea
Inequality \eqref{approx1} is sufficient to satisfy inequality \eqref{coshinus}.
\\

Then the approximate entropy with an island has the following form
\bea\label{approxent}
S_{\cI} = \frac{2\pi a^2}{G}+\frac{c}{6}\log \left(\frac{f(a)f(b)}{\kappa^4} \right)+\frac{c}{3} \kappa (r_\ast(b)-r_\ast(a))-\frac{2c}{3} e^{\kappa (r_*(a)-r_*(b))}.
\eea
The derivative of \eqref{approxent} with respect to $a=r_+ + X$,  $ X \ll r_+$ is expanded over $X$. In particular the following expansion is performed
\bea
\frac{1}{X+r_+-r_-}  \simeq \frac{1}{(r_+-r_-)} \left( 1-\frac{X}{r_+-r_-} \right),
\eea
that is true if the following inequality holds
\be\label{approx2}
{\mathbf Y}_2: \, \, X \ll r_+-r_-.
\ee
Then extremization of \eqref{approxent} with respect to $a$ gives
\be\label{xx}
X = \frac{c^2 G^2 r^2_+ (r_+ -r_-)^2
	e^{-\frac{(b-r_+) (r_+ -r_-)}{r^2_+}}
	\left(\frac{b-r_-}{r_+-r_-}\right)^{\frac{r^2_-}{r^2_+}}}{(b-r_+) \left[c G
	(r_+-2 r_-)+12 \pi  r^2_+ (r_- -r_+)\right]^2}.
\ee
Expanding of \eqref{xx} with respect to $G$ gives
\be\label{a}
X =  \frac{c^2 G^2 e^{2 \kappa (r_+-b)}
	\left(\frac{b-r_-}{r_+-r_-}\right)^{\frac{r^2_-}{r^2_+}}}{144 \pi
	^2 r^2_+ (b-r_+)} + O(G^3),
\ee
that is true (see denominator of \eqref{xx}) if the inequality holds
\be\label{approx3}
{\mathbf Y}_3: \, \,  \frac{c G |r_+-2 r_-|}{12 \pi  r^2_+} \ll r_+-r_-.
\ee
Two inequalities in \eqref{lta} can be rewritten as
\be\label{latetime}
t_b \gg \frac{1}{2} (b-r_+)+\frac{r^2_+}{2(r_+-r_-)} \log{\left[\frac{b-r_+}{X} \left( \frac{X+r_+-r_-}{b-r_-} \right)^{\frac{r^2_-}{r^2_+}} \right]}, \quad t_b \gg \frac{2 r^2_+}{r_+-r_-}.
\ee
The right sides of the inequalities \eqref{latetime} diverge in the extremal case, but in near-extremal case they are limited from below at least due to \eqref{approx3}.
\\

To control the degree of  smallness of the left sides of \eqref{approx1}, \eqref{approx2}, \eqref{approx3} we introduce 
\bea
\label{apprr1rho}
Y_1(\rho) &=& \rho \, \, 2 e^{\kappa (r_*(a)-r_*(b))}-1 ,\\\label{apprr2rho}
Y_2(\rho) &=& \rho \, \, X - r_++r_-, \\\label{apprr3rho}
Y_3(\rho) &=& \rho \, \, \frac{c G |r_+-2 r_-|}{12 \pi  r^2_+}- r_++r_-.
\eea
We take some particular  $\rho>1$, for instance $\rho=5$, and see when $Y_{1,2,3} (\rho)<0$.
The quantities \eqref{apprr1rho}, \eqref{apprr2rho}, \eqref{apprr3rho} under the constraint equation \eqref{stateequation} are
\bea
Y_1(\rho)\Big|_{\text{on constr.}} &=& \rho \, \,  2 e^{\kappa (G M (1+\alpha) + X-b)} \left( \frac{X}{b-G M (1+\alpha)} \right)^{\frac{1}{2}} \left( \frac{b-G M (1-\alpha)}{X+2 \alpha GM} \right)^{\frac{(1-\alpha)^2}{2 (1+\alpha)^2}}-1, \nn \\\label{apprr1rhoconstraint}\\ \label{apprr2rhoconstraint}
Y_2(\rho)\Big|_{\text{on constr.}} &=& \rho \, \, X - 2 \alpha G M, \\\label{apprr3rhoconstraint}
Y_3(\rho)\Big|_{\text{on constr.}} &=& \rho \, \, \frac{c (1-3\alpha)}{12 \pi  M (1+\alpha)^2}- 2 \alpha G M,
\eea
where $X$ \eqref{a} under constraint equation \eqref{stateequation} is
\be
X = \frac{c^2  e^{2 \kappa (G M (1+\alpha)-b)}
	\left(\frac{b-G M (1-\alpha)}{2 \alpha GM}\right)^{\frac{(1-\alpha)^2}{(1+\alpha)^2}}}{144 \pi
	^2 M^2 (1+\alpha)^2 (b-GM (1+\alpha))}.
\ee

\begin{figure}[h!]
\begin{center}
	\includegraphics[width=42mm]{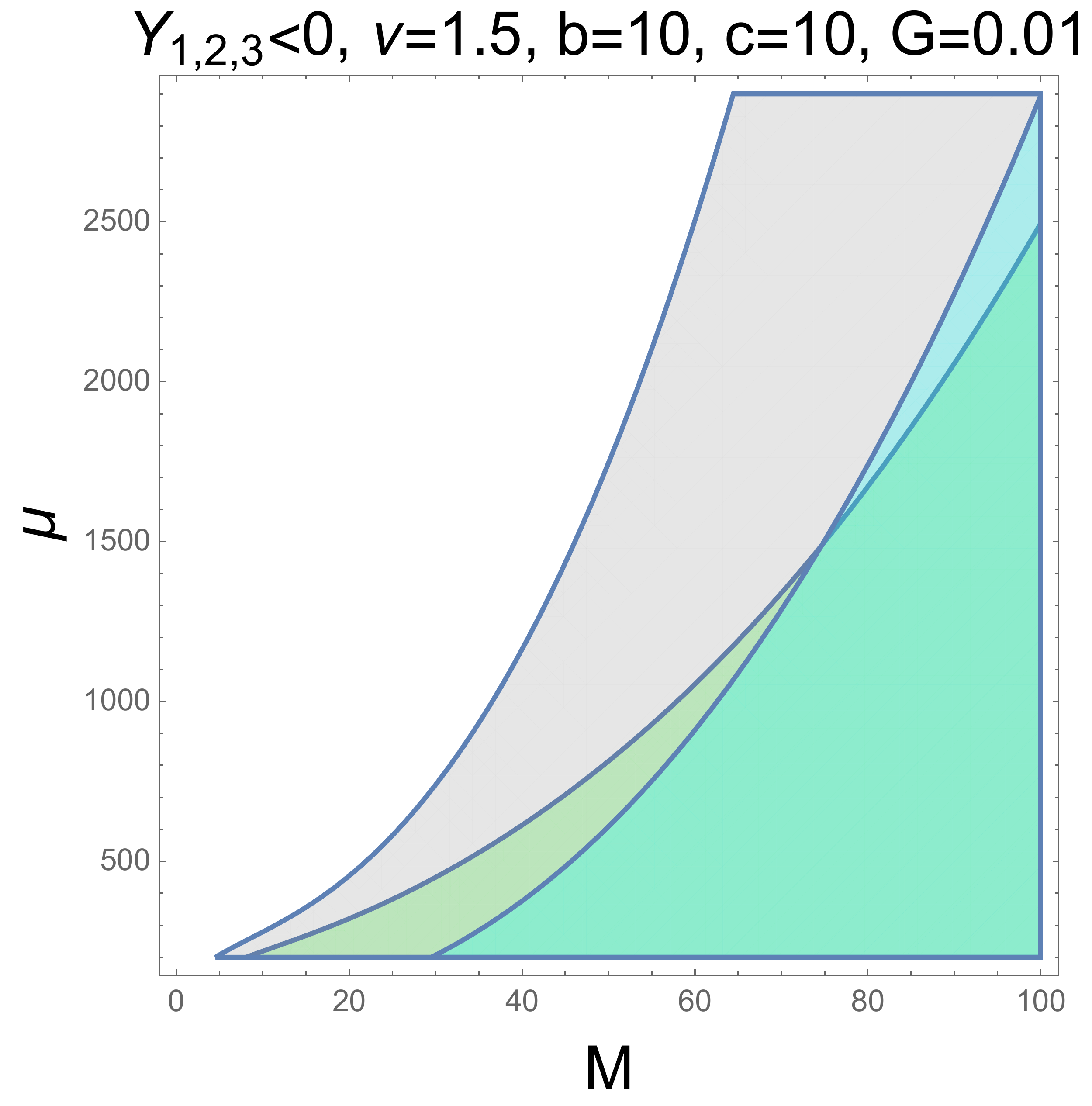}
	\qquad \includegraphics[width=44mm]{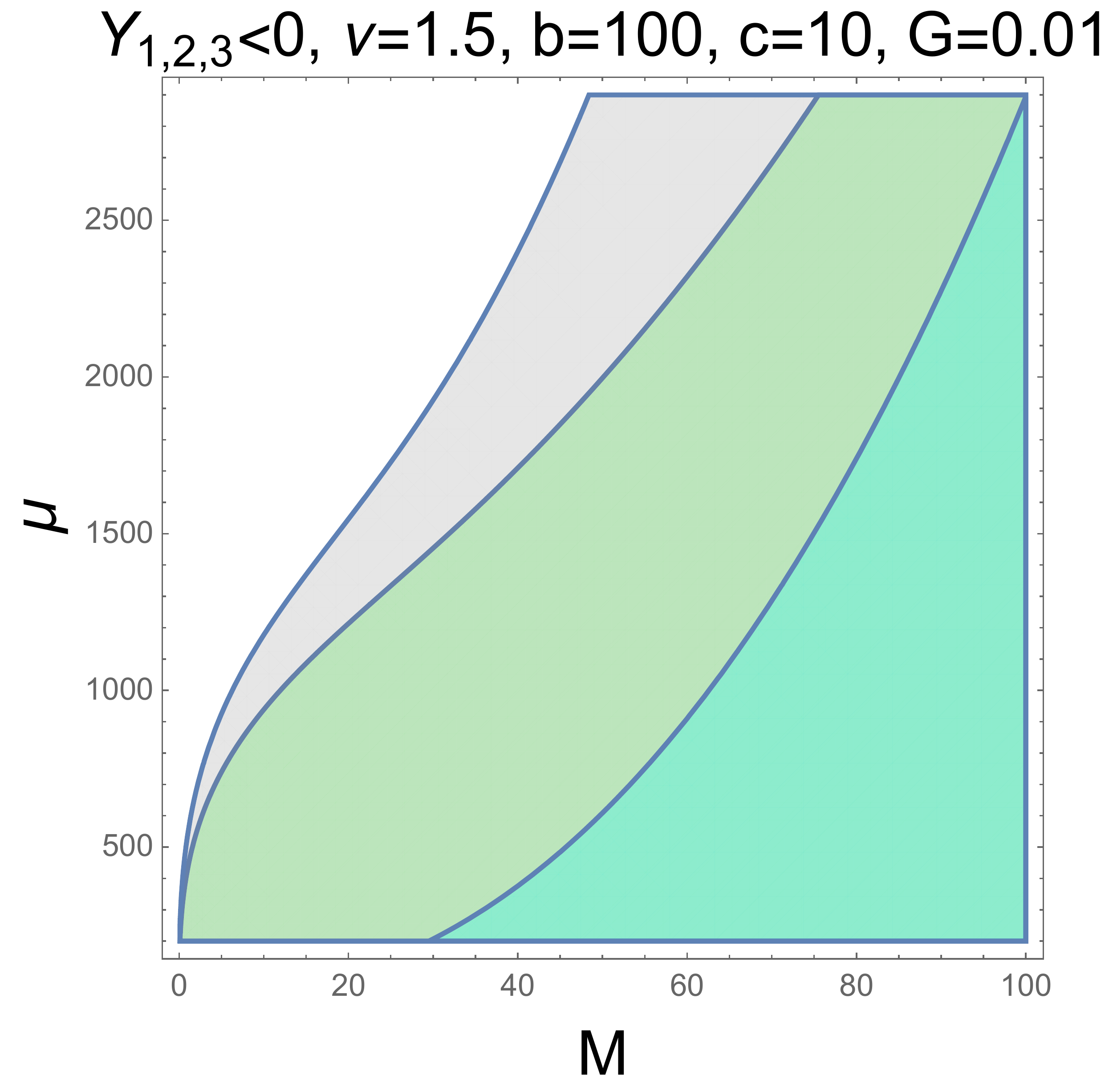}
	\qquad \includegraphics[width=46mm]{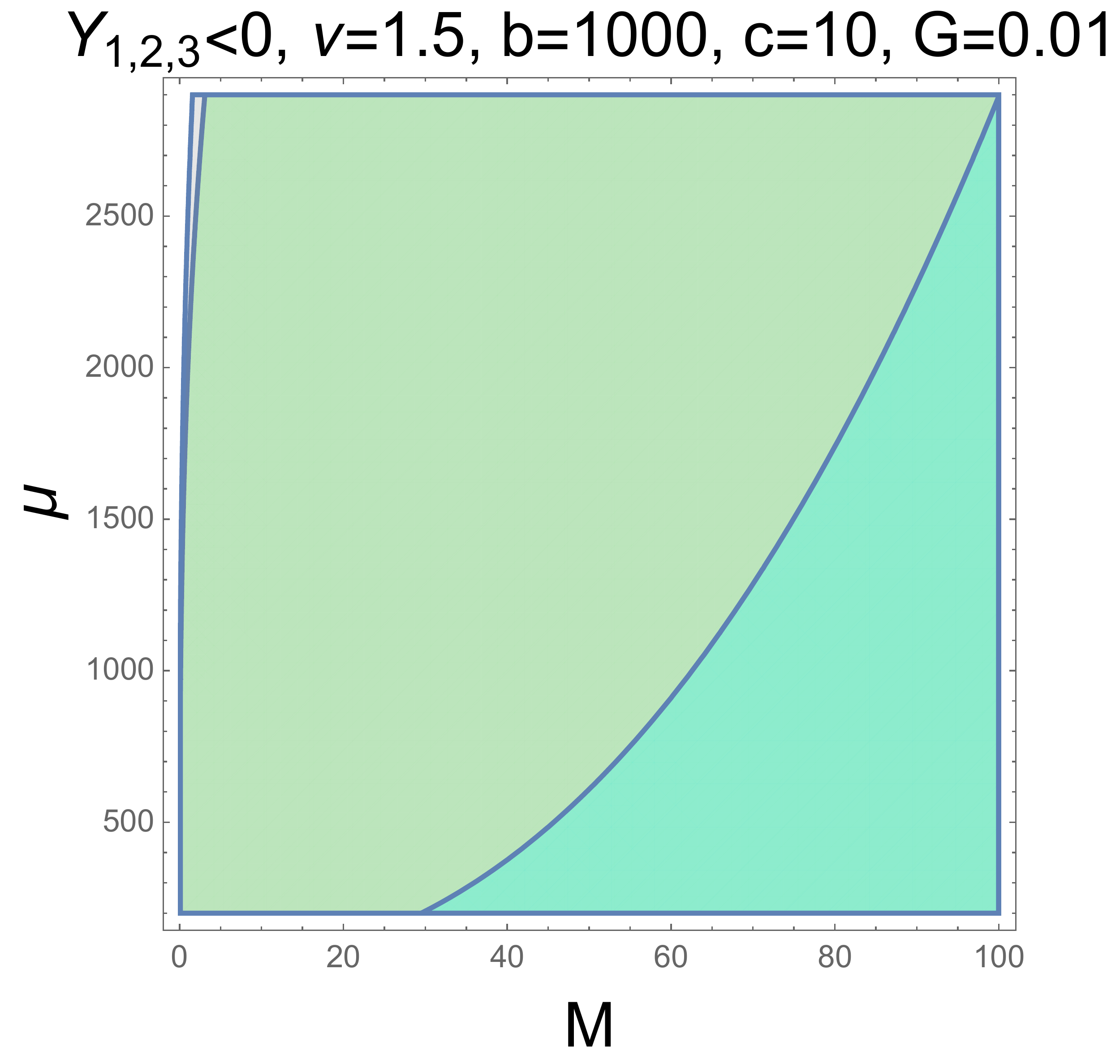}\\   A) \hspace{50mm}B) \hspace{50mm} C)\\\,\\
    \includegraphics[width=42mm]{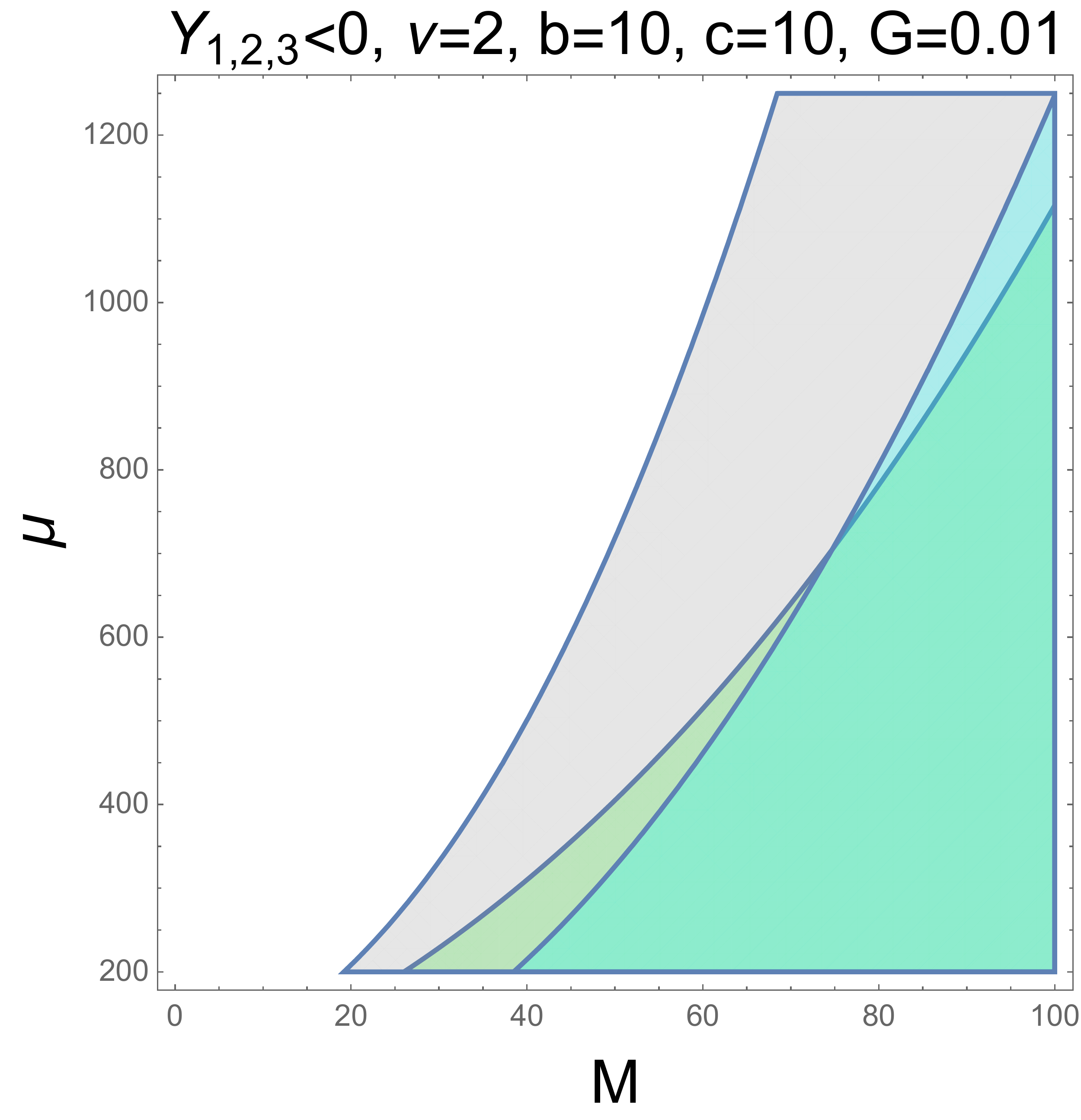}
	\qquad \includegraphics[width=44mm]{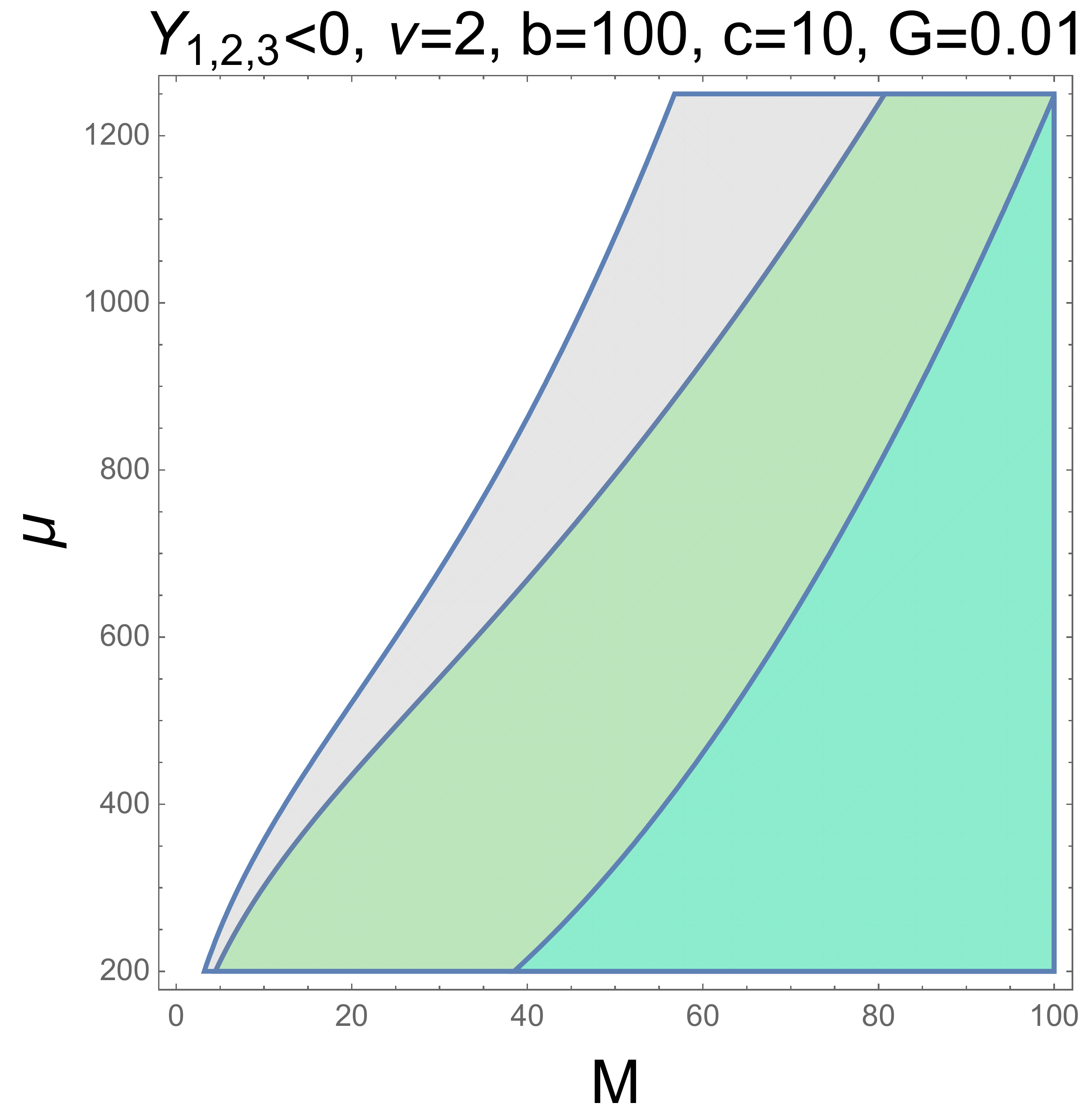}
	\qquad \includegraphics[width=46mm]{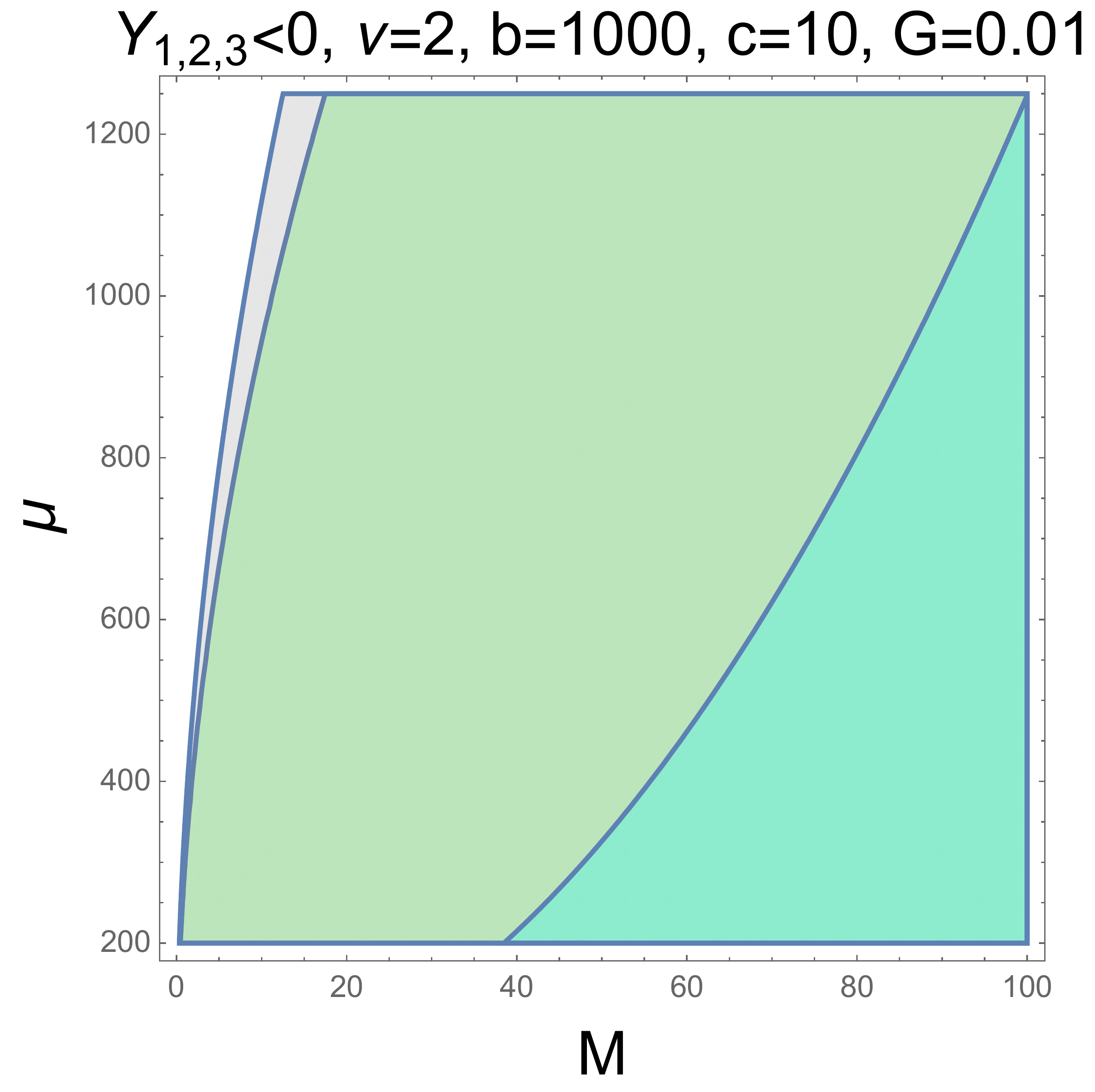}\\   D) \hspace{50mm}E) \hspace{50mm} F)\\
	\caption{Numerical analysis of inequalities $Y_{1,2,3} (\rho)<0$ on the  $(M,\mu)$ plane  for $\nu=1.5$ and $\nu=2$,  various $b= 10, 100, 1000$ and $\rho = 5$. The green regions correspond to $Y_1 < 0$, the gray regions correspond to $Y_2 < 0$, the cyan regions correspond to $Y_3 < 0$.}%{\bf Math.file: Graphs-Draft-Charged-BH-withMu-IA.nb}}
	\label{fig-approx}
	\end{center}
\end{figure}

Numerical analysis of inequalities $Y_{1,2,3}(\rho)<0$ \eqref{apprr1rhoconstraint}, \eqref{apprr2rhoconstraint}, \eqref{apprr3rhoconstraint} under the constraint equation \eqref{stateequation} on the  $(M,\mu)$ plane  for $\nu=1.5$ and $\nu=2$, various $b= 10, 100, 1000$
and $\rho = 5$ is presented in Fig.\ref{fig-approx}. It can be seen that the condition $Y_3<0$ 
	is the strongest in most cases, except for relatively small $b$ and big $\mu$, as shown in Fig.\ref{fig-approx}.A and Fig.\ref{fig-approx}.D. The larger the $\nu$ and $\mu$, the larger the value of the critical mass $M_{cr}$, up to which $M>M_{cr}$ the evaporation process can be considered. We also see that there is practically no dependence of $M_{cr}$ on $b$.
\\

Under the constraint equation \eqref{stateequation} the inequality ${\mathbf Y}_3$ \eqref{approx3} can be written as
\bea\label{approx3constraint}
{\mathbf Y}_3: \, \,   \frac{c (1-3\alpha)}{12 \pi  M (1+\alpha)^2} \ll 2 \alpha G M,\eea
that for small $\alpha$ can be simplified as 
\bea\label{Y3-simple}
{\mathbf Y}_3: \, \,  \frac{c}{24 \pi G M^2} \ll \alpha.
\eea
$$\,$$

Summarizing the above consideration we can say that approximations that provide validity of representation \eqref{islandch} for not too small $b$ are simply reduced to one inequality \eqref{Y3-simple}.
Taking into account the requirement to be close to the extremal case  $\alpha\ll1$, we get just the inequality
\be\label{maincond}
 \frac{c}{24 \pi G M^2}  \ll \alpha \ll 1.
\ee

Note that as can be seen from \eqref{maincond} that at sufficiently small $\alpha$ (large $\mu$ and/or $\nu$) and large $c$ the inequality \eqref{maincond} may not be satisfied.

\end{document}